\documentclass[twocolumn]{aastex6} 
\usepackage{amsmath} 
\usepackage{amssymb} 
\usepackage{amsfonts}
\usepackage{mathrsfs}
\usepackage{graphicx}

\shortauthors{Roth \& Kasen}
\shorttitle{What sets TDE line profiles?}

\begin{document}

\title{What sets the line profiles in tidal disruption events?}

\author{Nathaniel Roth\altaffilmark{1}}
\author{Daniel Kasen\altaffilmark{2,3,4}}
\email{nroth@astro.umd.edu}
\altaffiltext{1}{Department of Astronomy and Joint Space-Science Institute, University of Maryland, College Park, MD 20742, USA}
\altaffiltext{2}{Department of Physics, University of California, Berkeley, CA 94720, USA}
\altaffiltext{3}{Department of Astronomy and Theoretical Astrophysics Center,
University of California, Berkeley, CA 94720, USA}
\altaffiltext{4}{Nuclear Science Division, Lawrence Berkeley National Laboratory,
Berkeley, CA 94720, USA}

\begin{abstract} 
We investigate line formation in gas that is outflowing and optically thick to electron scattering, as may be expected following the tidal disruption of a star by a supermassive black hole. Using radiative transfer calculations, we show that the optical line profiles produced by expanding TDE outflows most likely are primarily emission features, rather than the P-Cygni profiles seen in most supernova spectra. This is a result of the high line excitation temperatures in the highly irradiated TDE gas.  The outflow kinematics cause the emission peak to be blueshifted and to have an asymmetric red wing. Such features have been observed in some TDE spectra, and we propose that these may be  signatures of outflows. We also show that non-coherent  scattering off of hot electrons can broaden  the emission lines by  $\sim 10,000$~km~s$^{-1}$, such that the line width in some TDEs may be set by the electron scattering optical depth rather than the gas kinematics. The scattering broadened line profiles produce distinct, wing-shaped profiles that are similar to those observed in some TDE spectra. The narrowing of the emission lines over time in these observed events may be related to a drop in density rather than a drop in line-of-sight velocity.
\end{abstract}


\section{Introduction}
Wide-field surveys are discovering a growing number of TDEs at optical wavelengths \citep{van-Velzen2011,Gezari2012,Arcavi2014,Holoien2014,Holoien2016-1,Blagorodnova2017,Hung2017}. Despite the abundance of data, the origin of the optical emission remains unclear.  Observations suggest that the optical radiation is produced from a region many times larger than the tidal disruption radius. A number of explanations have been proposed to explain this extended emission region:  the formation of a quasi-static reprocessing envelope surrounding an accretion disk \citep{Loeb1997,Bogdanovic2004,Guillochon2014}, the ejection of a quasi-spherical outflow of stellar debris \citep{Strubbe2009,Lodato2011,Miller2015,Metzger2016}, or collisions occurring in the stellar debris streams falling back to the black hole \citep{Kochanek1994,Kim1999,Piran2015,Krolik2016,Jiang2016-1,Bonnerot2017-1}. A number of TDE-impostor scenarios have also been proposed. The optical emission might result from a flash of high-temperature radiation that illuminates pre-existing broad-line region clouds, perhaps as a result of an extreme supernova embedded in dense nuclear gas or a separate type of supermassive black hole accretion flare \citep{Saxton2018}. Alternatively, grazing collisions of stars undergoing steady mass transfer with the black hole might give rise to optical flares that can be confused for TDEs \citep{Metzger2017}.

Optical and UV spectra, when available, can provide a wealth of information about the kinematics and conditions in the emitting gas that may allow us to test these hypotheses. However, interpreting the spectra is challenging because of the high gas densities and uncertain geometry \citep{Gaskell2014,Guillochon2014,Strubbe2015}. A previous study \citep[][hereafter R16]{Roth2016} presented detailed radiative transfer calculations of the spectra from an optically thick, spherical TDE envelope. These calculations assumed that electron scattering was coherent, and that the motion of the envelope gas was turbulent rather than outflowing. The simplifications prevented an analysis of how the kinematics and thermal properties of the line-emitting gas shape the line profiles.

A number of fundamental questions regarding the spectra of TDEs remain to be addressed:

1) Many models of the optical emission for TDEs argue that the optical emission is produced in a nearly spherical outflow of  stellar debris ejected with velocities $\sim 10^4$~km~s$^{-1}$. Indeed, radio detections of some TDE provide evidence of non-relativistic outflows \citep{Alexander2016,Alexander2017} \citep[but see][for alternative interpretations]{van-Velzen2016-1,Krolik2016}. In analogy to supernovae, one might expect such outflows to produce P-Cygni spectral line profiles, with both absorption and emission components. In contrast, TDEs generally show pure emission profiles, with the possible exceptions of PTF-11af \citep{Chornock2014}, iPTF-16fnl \citep{Brown2018}, and ASASSN-15lh \citep{Dong2016,Leloudas2016,Margutti2017} (although for the latter's TDE identification is debated). It is thus unclear whether the line profiles of TDEs are consistent with outflows, and if so, why TDE spectra differ qualitatively from those of supernovae. 

2) While the emission lines in TDEs generally have widths corresponding to velocities of $\sim 10^4$~km~s$^{-1}$, there is a large spread in widths between individual events \citep[Figure 15 of][]{Hung2017}. In some cases, the line widths narrow over time while the luminosity drops \citep{Holoien2014,Holoien2016-1,Holoien2016-2,Blagorodnova2017,Brown2018}, which is opposite to the behavior seen in reverberation mapping of quasars \citep{Holoien2016-2}. It is unclear how the line widths in TDEs are related to the kinematics or other physics  in the emitting gas.

3) In a number of TDEs, the early spectra show emission lines with blueshifted centroids \citep{Holoien2014,Arcavi2014,Holoien2016-2}. There is also  evidence of a line asymmetry in these events, with the emission extending farther to the red side of the line than to the blue. In all cases when multi-epoch spectroscopy is available, the centroid moves closer to the host rest frame over time, and the line becomes more symmetric. One potential explanation \citep{Gezari2015, Brown2018} for the blueshifted component of the \ion{He}{2} $\lambda4686$ line is that it is a blend of \ion{C}{3} and \ion{N}{3} emission (Bowen fluorescence). While this seems like a promising explanation in some cases, it cannot explain blueshifts of H$\alpha$, and it may not provide blue enough emission to match some of the observations.

In this paper, we use radiative transfer models to study the fundamental physics of TDE line formation and address the above questions. We find that optically thick TDE outflows probably produce H and He lines  with pure emission (not P-Cygni) profiles, similar to what is seen in observed spectra. The lack of line absorption is a result of the high line source function realized in the strongly irradiated TDE gas. We show that typical signatures of an outflow include a blueshifted line-emission peak, and an asymmetric, extended red wing --  both features that have been seen in the spectra of  observed TDEs. In addition, we show  that the  line widths and their time evolution can be significantly affected by non-coherent (Compton) electron scattering; thus, the line widths in some TDEs may be related to thermal broadening in addition to (or rather than) the bulk kinematics of the line-emitting gas. 

In Section~\ref{sec:ModelSetup}, we describe the basic features of our model setup. We then consider simple and parameterized calculations. to provide insight into the physics shaping line formation in TDEs.  Section~\ref{sec:TexVariation} demonstrates how the line source function controls whether or not a P-Cygni or pure emission line profile will be generated in an outflow. Section~\ref{sec:ScatterBroadening} illustrates how repeated non-coherent scattering of photons from thermal electrons can broaden an emission line, and applies this idea to fit the line profiles of ASASSN-14li. In Section~\ref{sec:CombinedCalculations} we move on to more detailed calculations that combine the aforementioned effects. We explore how H$\alpha$ lines may be shaped by the hydrodynamic parameters, and present a fit to ASASSN-14ae. We also discuss how the \ion{He}{2} to H$\alpha$ line strengths should evolve. We summarize our conclusions and characterize the limitations of our work in Section~\ref{Sec:Discussion}. 

\section{Model Setup}
\label{sec:ModelSetup}

We perform radiative transfer calculations in spherical symmetry by using an adapted version of {\tt Sedona}, a Monte Carlo radiative transfer code \citep{Kasen2006, Roth2015}. We consider gas that is distributed with a density profile $\rho(r) \propto r^{-2}$ between an inner radius $r_{\rm in}$ and outer radius $R$. The gas is outflowing, with a velocity profile $v(r)$ specified below. We take the gas properties to be time-independent, which is appropriate for the regime where the radiation diffusion time is small compared to the timescale over which the gas properties change. More detailed calculations would solve for the time-dependent hydrodynamics of the envelope density and velocity structure, along with the consequent time-dependent radiation properties.
 
We model  spectra in a wavelength interval surrounding a single line of interest, in this case the H$\alpha$ transition in hydrogen (though the qualitative features of line formation we discuss probably apply to other optical lines of interest).  In the Monte Carlo calculation, continuum photon packets are emitted from an inner, absorbing boundary at radius $r_{\rm th}$, representing the radius of thermalization of the continuum. We specify the luminosity of the continuum emission at $r_{\rm th}$, and take the continuum spectrum to be flat (constant in $F_\lambda$), given the narrow wavelength interval modeled. Photon packets are further emitted and absorbed at each location within the gas, in accordance with the line opacity and source function. In our simplified calculations, we set these two quantities  parametrically, while in our more detailed models we derive them by solving the non-LTE rate equations for the ionization and excitation states of the gas under statistical equilibrium (see Appendix~\ref{sec:Method}). We tally the escaping packets from both the line and the continuum to generate the observed spectrum.

\section{Why Not a P-Cygni Profile? The Role of the Line Source Function}
\label{sec:TexVariation}

We first consider simple models that help explain TDE line profiles. We consider an expanding outflow with a homologous velocity profile
\begin{equation}
v(r) =  v_{\rm sc}\, \frac{r}{R} \, \, ,
\end{equation}
where $v_{\rm sc}$ is the velocity at the outer boundary $R$. For an atomic line with intrinsic width given by the Doppler velocity $v_{\rm D}$, photons interact with the gas in the so-called resonance region, which for a homologous outflow has a physical size
\begin{equation}
\Delta r = v_{\rm D} \left(\frac{dv}{dr}\right)^{-1}  = \frac{v_{\rm D}}{v_{\rm sc}} R \, \, .
\end{equation}
In outflows with strong velocity gradients ($v_{\rm sc} \gg v_{\rm D}$),  $\Delta r$ is small compared to the scales over which the gas properties vary, and the Sobolev approximation \citep{Sobolev1960} can be used to calculate the resulting line profile \citep[e.g.][]{Jeffery1990}. 

The optical depth integrated across the resonance region at some radius $r$ is a local quantity, $\tau_s(r)$, called the Sobolev optical depth. The profile of a line is determined  by $\tau_s(r)$ and the line source function, given by
\begin{equation}
S_{\lambda} = \frac{2 h c^2}{\lambda^5} \frac{1}{\exp \left( \frac{h c }{\lambda_{\rm line} k_B T_{\rm ex}}\right) - 1} \, \, ,
\label{eq:LineSourceFunction}
\end{equation}
 where $T_{\rm ex}$ is referred to as the line excitation temperature \citep{Jefferies1958}. Here, $T_{\rm ex}$ only corresponds to a true thermodynamic temperature when the atomic level populations are in local thermodynamic equilibrium; more generally,
  \begin{equation}
 \exp \left( \frac{h c }{\lambda_{\rm line} k_B T_{\rm ex}}\right)  = \frac{g_2}{g_1} \frac{n_1}{n_2} \, \, ,
 \end{equation}
 where  $n_1, n_2$ are the level populations and $g_1, g_2$  the statistical weights of the lower and upper levels (respectively) of
 the atomic transition.
 
\begin{figure}[htb!]
\includegraphics[width =0.5\textwidth]{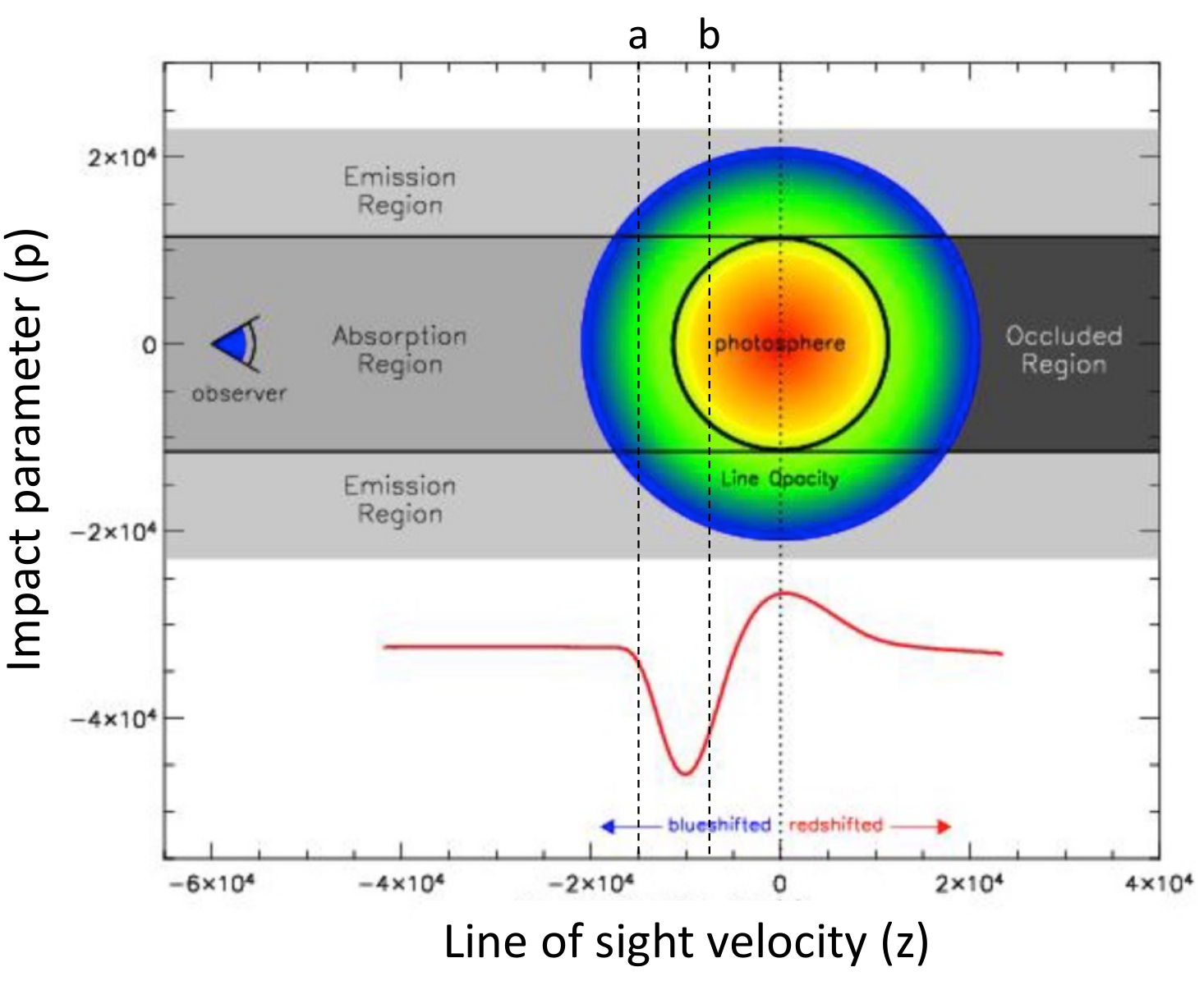}
\caption{Schematic of line formation in an expanding atmosphere. For TDE atmospheres, the line source function can be large enough that the gas in the ``absorption region" emits more than it absorbs. In this case, the line will produce a pure emission profile.}
\label{fig:PCygniCartoon}
\end{figure}

 Line formation in rapidly expanding outflows is familiar from studies of supernovae and stellar winds, 
and is illustrated in the schematic Figure~\ref{fig:PCygniCartoon}. In the simple, heuristic picture,  continuum flux is emitted from the surface of a sharp photosphere into tenuous, line-forming gas. The gas on the sides of the photosphere produces a line emission feature that peaks at the line center wavelength. The gas in front of the photosphere -- which is moving toward the observer -- obscures the continuum flux and produces a blueshifted absorption feature in the classic P-Cygni profile.  

The absorption component of the  P-Cygni profile shown in  Figure~\ref{fig:PCygniCartoon} is only produced when the gas in the ``absorption region" absorbs more than it emits. This occurs when $T_{\rm ex}$ is less than the brightness temperature, $T_{\rm bb}$, of the photosphere. In the common assumption of resonance line scattering (where  every line absorption  is  immediately followed by emission via the same atomic transition), the source function equals the mean intensity of the local radiation field, and $T_{\rm ex} < T_{\rm bb}$ due to the geometrical dilution of the continuum radiation emergent from the photosphere.

In highly irradiated TDE outflows, however, it is possible for $T_{\rm  ex}$ to deviate from the resonant scattering value. A self-consistent calculation would require that we simultaneously solve the non-LTE rate equations coupled with the radiative transfer equation. The former determine the emissivity and opacity of each line, while the latter determines quantities such as photoionization rates and mean radiative intensities at line wavelengths $\bar{J}_{\lambda,\rm line}$, which go into the non-LTE equations that determine the line emissivities and opacities.

Here, to illustrate the diversity of line profiles, we present transport models that ignore electron scattering and use  a simple parameterization for $T_{\rm ex}$. We  choose $T_{\rm ex}$ to vary linearly with the gas column density such that it is equal to some specified $T_{\rm ex,\,\,out}$ at $R$ and 10 times that value at $r_{\rm th}$.  Such behavior is consistent with the line source function we find in more detailed NLTE calculations (see  Section~\ref{sec:CombinedCalculations}). 
 
 As a fiducial model, we choose  $r_{\rm in} = 10^{14}$~cm, $R = 10^{15}$~cm, $v_{\rm sc} = 10^{4}$ km s$^{-1}$, and an envelope mass of $0.25$~$M_{\odot}$ (which gives $\rho(r_{\rm in}) = 4.32 \times 10^{-12}$~g~cm$^{-3}$). We set $v_D = 400$~km~s$^{-1}$, $r_{\rm th} = 3 \times 10^{14}$~cm, and $\kappa = 0.03$~cm$^{2}$~g$^{-1}$, in this case constant with radius. For the source function described above, the choice for the absolute strength of the continuum flux affects the emission or absorption properties of the line. We choose to set it so that the specific luminosity is $2.5 \times 10^{38}$~erg~s$^{-1}$~$\AA^{-1}$ at all wavelengths, which corresponds to $T_{\rm bb} = 2.5 \times 10^4$~K at line center.

\begin{figure}[htb!]
\includegraphics[width =0.5\textwidth]{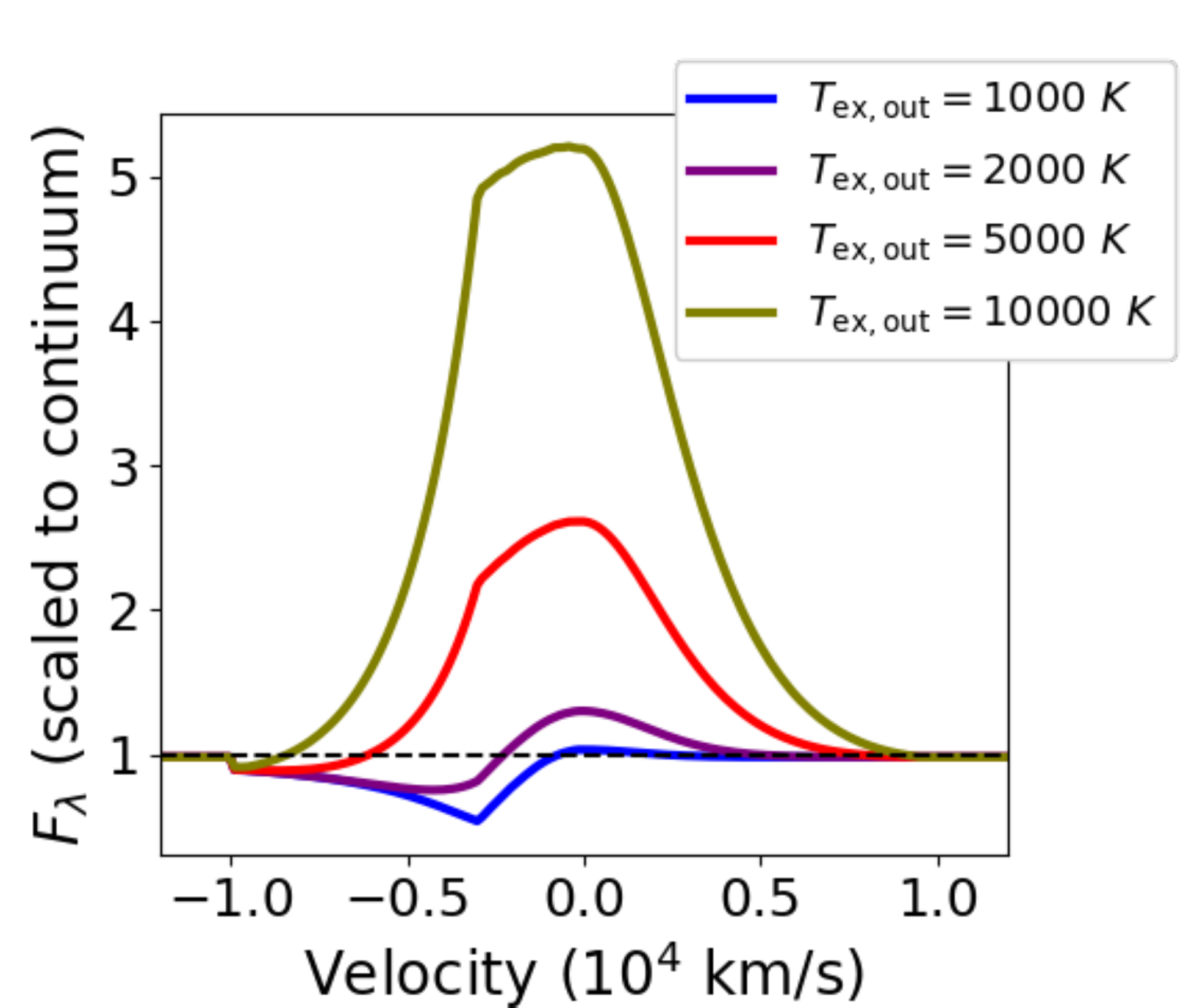}
\caption{A demonstration of how the line excitation temperature $T_{\rm ex}$ (a measure of the line source function) affects line profiles in an expanding atmosphere, in the simplified case of no electron scattering. Here, $T_{\rm ex}$ is set in simple parameterized fashion, ranging from the specified value at the outer radius to 10 times that value at the inner photosphere for the continuum. A classical P-Cygni profile, with blueshifted absorption and redshifted emission tail, is generated for an intermediate value of $T_{\rm ex}$. For lower values of $T_{\rm ex}$, the profile displays prominent blueshifted absorption with little corresponding emission. For higher values of $T_{\rm ex}$, the line  appears entirely in emission. In this figure the maximum velocity of the gas is $v_{\rm sc} = 10^4$~km~s$^{-1}$ , and the  velocity at the continuum thermalization depth is $v_{\rm th} = -3 \times 10^3$~km~s$^{-1}$. }
\label{fig:VaryExcitationTemperature}
\end{figure}

The resulting model line profiles are shown in Figure~\ref{fig:VaryExcitationTemperature}. For $T_{\rm ex, \,\, out}$ = 2000~K, the line profile is very similar to that of resonant scattering. Reducing $T_{\rm ex, out}$ to 1000~K  produces a line with prominent blueshifted absorption and very little emission. Raising $T_{\rm ex, out}$ to 5000~K makes the blueshifted absorption shallower and extended over a smaller range of wavelengths.

For $T_{\rm ex, out}$ = 10,000~K, the line appears entirely in emission. It turns out that this choice is close to the line source function  we compute in Section~\ref{sec:CombinedCalculations}, following the more detailed NLTE procedure described in Appendix~\ref{sec:Method}. The high values of $T_{\rm ex}$ arise in the full calculation because of the high radiative luminosity emanating from the inner TDE engine, the limited spatial extent of the line-emitting gas, and the high scattering depth that traps the radiation and raises its mean intensity compared to the free-streaming case.

For homologous expansion, all emission and absorption at a given wavelength corresponds to gas on a $z = $ constant plane (see coordinate system labels on the Figure~\ref{fig:PCygniCartoon}). For $|z| > r_{\rm th}$ on the approaching side, the plane fully covers the continuum photosphere, as demonstrated by the plane labeled `a' in Figure~\ref{fig:PCygniCartoon}, whereas for $|z| < r_{\rm th}$ the plane cuts through the photosphere so that only its edges are blocked (as in plane `b'). Therefore, we see a line feature at the approaching velocity of the photosphere
 $v_{\rm th} \equiv  - v_{\rm sc} r_{\rm th}/R$, which can correspond to the point of maximum absorption for sufficiently low $T_{\rm ex}$, or a shoulder in the emission for sufficiently high $T_{\rm ex}$.

While P-Cygni profiles were not generated for the H$\alpha$ and \ion{He}{2} line profiles calculated below, other lines, such as those from highly ionized carbon and nitrogen, would potentially display blueshifted absorption in the same environment, as has sometimes been seen in the UV spectra of TDEs \citep{Chornock2014,Blagorodnova2017}, where the line source functions may be closer to resonant scattering. 

\section{The role of Non-coherent Electron Scattering in Setting Line Widths and Line-narrowing}
\label{sec:ScatterBroadening}

In addition to the kinematic effects just discussed, spectral lines can be broadened by multiple scatterings of photons by electrons \citep{Dirac1925, Munch1948, Chandrasekhar1950}. 
 Electrons in random thermal motion have velocities $v_e \approx \sqrt{k_B T_e/m_e}$, where $T_e$ and $m_e$ respectively are the electron temperature and mass. Photons with small energies, as compared to the electron rest energy, pick up Doppler shift factors of order $v_e/c$ in each scattering event. After $N$ scattering events, a photon has undergone an effective diffusion process in wavelength space such that the line photon broadens by a factor of $\sim \sqrt{N} v_e/c$ (note that this behavior changes for large enough $N$ and large enough photon energy; see Appendix~\ref{Comptonization}). Astrophysical examples of this type of line broadening include emission lines in Wolf-Rayet stars \citep{Munch1950, Castor1970, Hillier1984}, absorption lines in O and B stars \citep{Hummer1967},  emission lines from AGN \citep{Kaneko1968, Weymann1970, Kallman1986, Laor2006}, Fe K$\alpha$ emission in x-ray binaries \citep{Ross1979, George1991}, and some supernovae \citep{Chugai2001, Aldering2006, Dessart2009-1, Humphreys2012, Gal-Yam2014, Fransson2014, Borish2015, Dessart2015, Dessart2016, Huang2018}.

When scattering-broadening dominates in a plasma of moderate electron scattering optical depth $\tau_e$, the result is a narrow line core consisting of un-scattered photons, surrounded by a broad component referred to as the ``wings'' or ``pedestal.'' These wings have the potential to be misinterpreted as kinematic broadening, leading to overestimates of bulk velocities \citep[e.g.][]{Chugai2001, Dessart2009-1}. A similar issue may arise in the interpretation of P-Cygni profiles when $\tau_e$ is non-negligible \citep{Auer1972}, .

Here, we present radiative transfer calculations that include the physics of non-coherent electron scattering, as descrifbed in Appendix~\ref{Comptonization}.
We use the same gas density profile as described in the previous section. We do not attempt to model the continuum emission, but rather consider only line emission coming from $r_{\rm in} = 10^{14}$~cm that then scatters on its way out to the observer. 

\begin{figure}
  \includegraphics[width=0.5\textwidth]{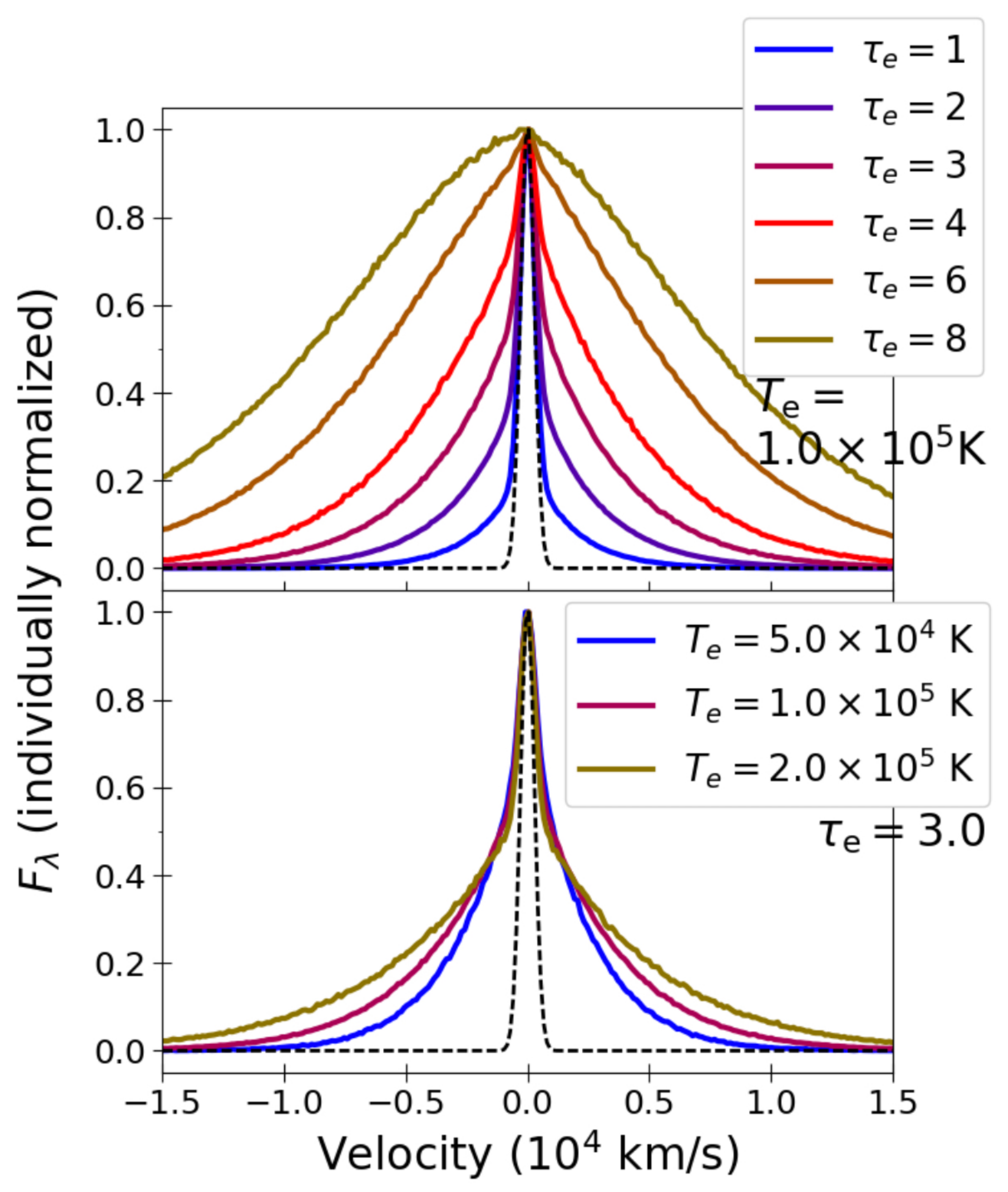}
  \caption{Demonstration of the line-broadening effect of non-coherent electron scattering. The top panel shows model line profiles for varying values of 
  electron scattering optical depth $\tau_e$ in a gas with no net velocity and  $T_e = 10^5$ K throughout. All fluxes have been scaled to the their value at line center. The characteristic core and wing profile is visible at lower values of $\tau_e$. As $\tau_e$ increases, the wings become wider and the core contribution becomes smaller. At sufficiently high $\tau_e$, the line appears broad without a distinguishable core. The bottom panel depcits model line profiles for varying values of $T_e$ with $\tau_e = 3$ throughout. The effect of increasing $T_e$ is mostly seen in the wings of the line.}
  \label{fig:LineBroadeningUnnormalized}
\end{figure}

Figure~\ref{fig:LineBroadeningUnnormalized} shows the resulting continuum-subtracted line profiles for different values of $\tau_e$ and $T_e$. For $\tau_e \lesssim 5$, the characteristic core-and-wing profile is visible, with a larger portion of the core escaping at lower optical depth \citep[cf.][]{Chugai2001}. The narrow core is composed primarily of line photons that have traveled all the way through the envelope without scattering. The wings are built up from photons that have diffused in frequency space as a result of multiple Doppler shifts from multiple electron scatterings. For larger optical depths ($\tau_e \gtrsim 5$) only the wings are visible. When we keep $\tau_e$ constant and vary $T_e$, we see that the wings of the line become broader, while the core of the line is mostly unaffected.

\begin{figure*}
    \includegraphics[width=\textwidth]{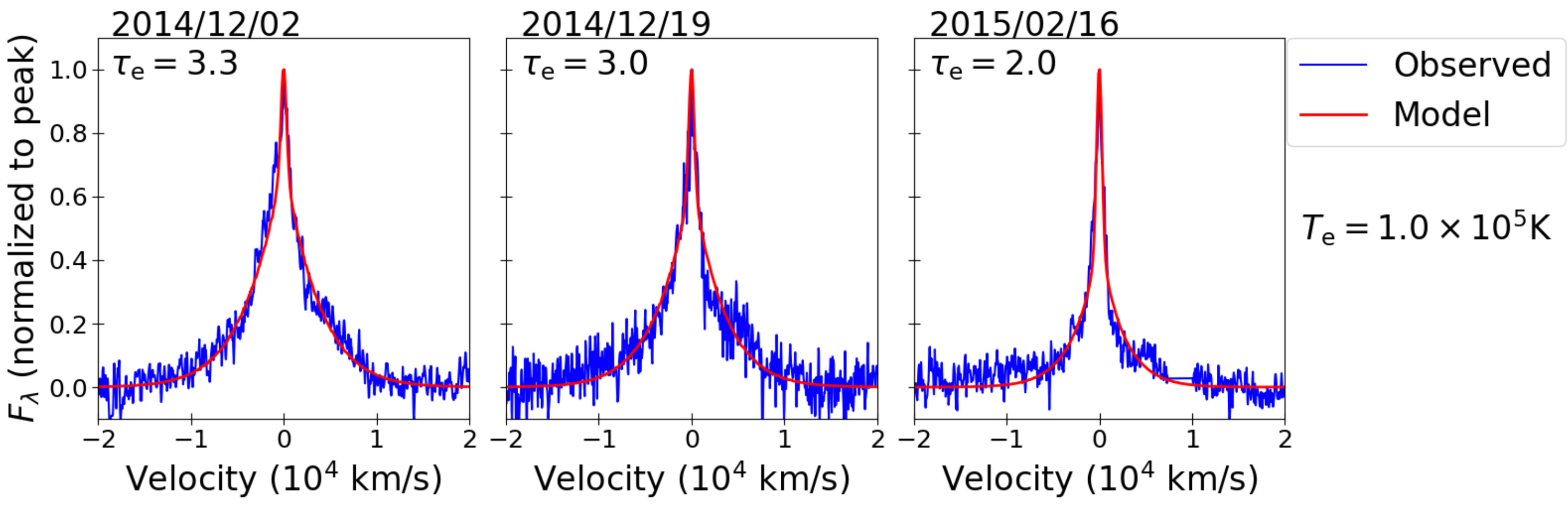}
  \caption{Model fits to the host-subtracted and continuum-subtracted H$\alpha$ line profiles of ASASSN-14li taken at three epochs \citep{Holoien2016-1}. The model accounts for non-coherent scattering of line photons by a layer of electrons at the specified optical depth $\tau_e$ and temperature $T_e$. The only parameter adjusted between the three fits is the value of $\tau_e$.}
  \label{fig:14liFits}
\end{figure*}

We compare these scatter-broadened line models to the host-subtracted spectra of the TDE ASASSN-14li \citep{Holoien2016-1}, for which we have subtracted a linear fit to the continuum near H$\alpha$. Model fits for three epochs are shown in Figure~\ref{fig:14liFits}. The value of $\tau_e$ was the single parameter that was changed to produce the three fits, with respective values of 3.3, 3.0, and 2.0 for the three selected epochs.

\begin{figure}
    \includegraphics[width=0.5\textwidth]{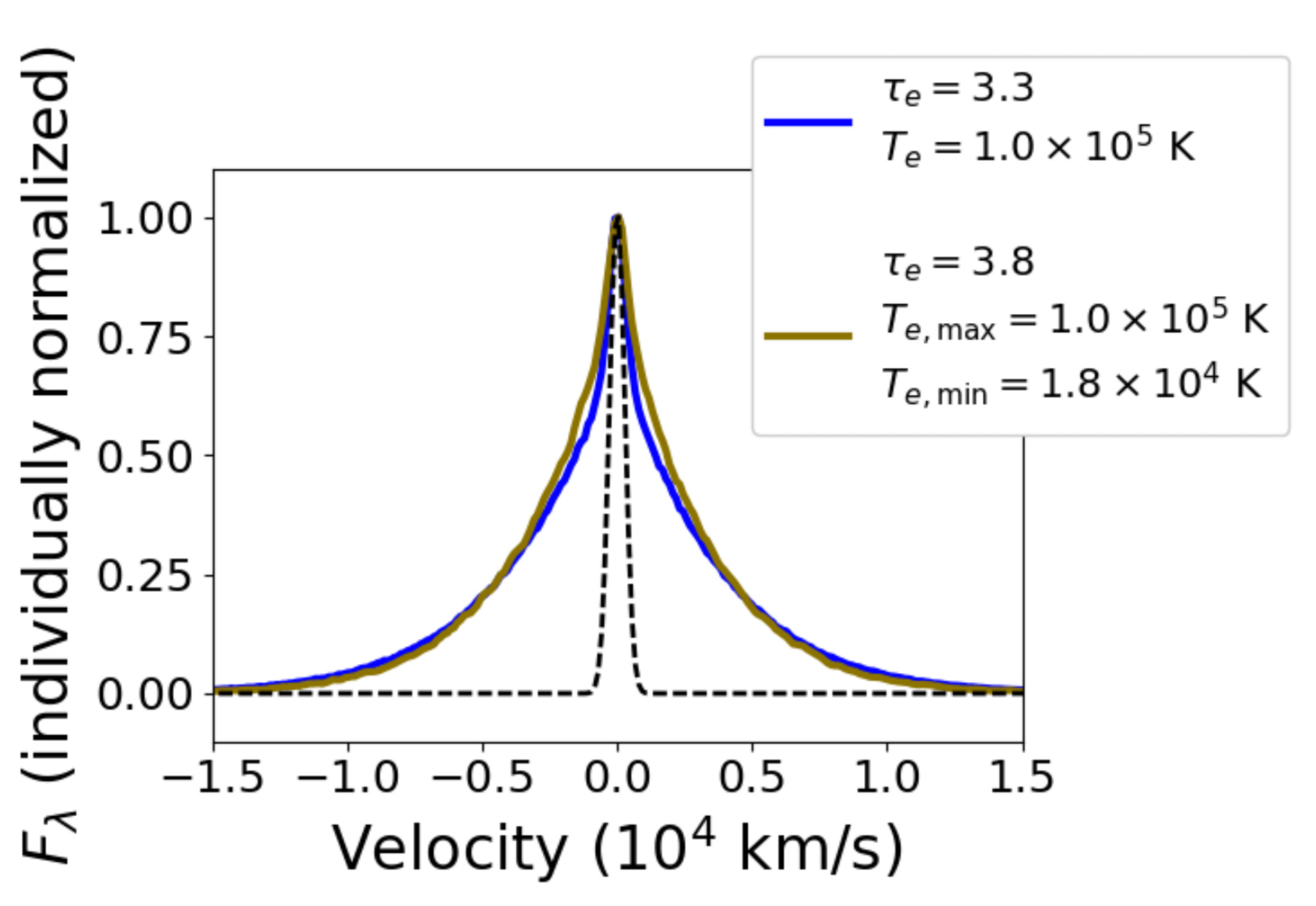}
  \caption{An illustration of the degeneracy between $\tau_e$ and $T_e$ in modeling the non-coherent electron scattering of line photons. The gold line profile corresponds to a model with $\tau_e = 3.8$ and a spatial gradient for $T_e$ in accordance with radiative diffusion, with a maximum $T_e = 10^5$~K and a minimum $T_e = 1.8 \times 10^4$~K. The resulting line profile is very similar to that produced with a constant $T_e = 10^5$~K and $\tau_e = 3.3$, which was used to fit the first line profile in Figure~\ref{fig:14liFits}.}
  \label{fig:14liTemperatureGradient}
\end{figure}

There exists some degeneracy between $\tau_e$ and $T_e$ when fitting line profiles using this model. This degeneracy is made more acute if we allow $T_e$ to vary with position, as we would expect in reality. To illustrate this, in Figure~\ref{fig:14liTemperatureGradient} we include a model in which $T_e$ follows an $r^{-3/4}$ relation. This corresponds to the diffusion approximation for the radiative energy density $e_{\rm rad}$ given our $r^{-2}$ density profile, with the added assumption that $T_e = \left(e_{\rm rad} / a_{\rm rad}\right)^{1/4}$. For $T_e = 10^5$~K at $r_{\rm in} = 10^{14}$ cm, this results in a temperature of $1.8 \times 10^4$~K at the outer radius $R = 10^{15}$~cm. For this temperature profile and for $\tau_e = 3.8$, the resulting line profile is very similar to the model used to fit the earliest epoch in Figure~\ref{fig:14liFits}, which used constant $T_e = 10^5$~K and $\tau_e = 3.8$. While the model that includes a temperature gradient is more realistic, the constant-temperature model achieves a fit of similar quality and only a modestly different value inferred for $\tau_e$.

The fitted values of $\tau_e$ should be considered lower bounds that roughly approximate the scattering optical depth above the thermalization depth of the line. In the model, the line photons were emitted at a constant radius and were not reabsorbed by either the line or by continuum processes. In order to obtain a similar line width in the more realistic case when photons are emitted at a range of radii, including close to the electron scattering photosphere, a higher optical depth will be required.

The high quality of the model fits to the line profiles of ASASSN-14li suggests that non-coherent electron scattering may have had a dominant effect in setting the line widths for this TDE. This would imply that the evolution of the line widths mostly reflects a reduction in optical depth over time, rather than kinematic behavior. 

\section{Calculations Combining Outflows and Electron Scattering}
\label{sec:CombinedCalculations}

In the previous sections, we used simplified setups to illustrate how outflows, the line source function, and non-coherent electron scattering (NCES) affect the line profiles. We now present more realistic calculations of line formation in TDE outflows that include NCES, along with the line source function and opacity derived from a more detailed NLTE analysis, which includes the effect of adiabatic reprocessing of the continuum (see Appendix~\ref{sec:Method} for details). The procedure we use can be applied to any line, but we use H$\alpha$ for concreteness.

\subsection{Homologous Expansion at Different Maximum Velocities}

\begin{figure}[htb!]
\includegraphics[width =0.5\textwidth]{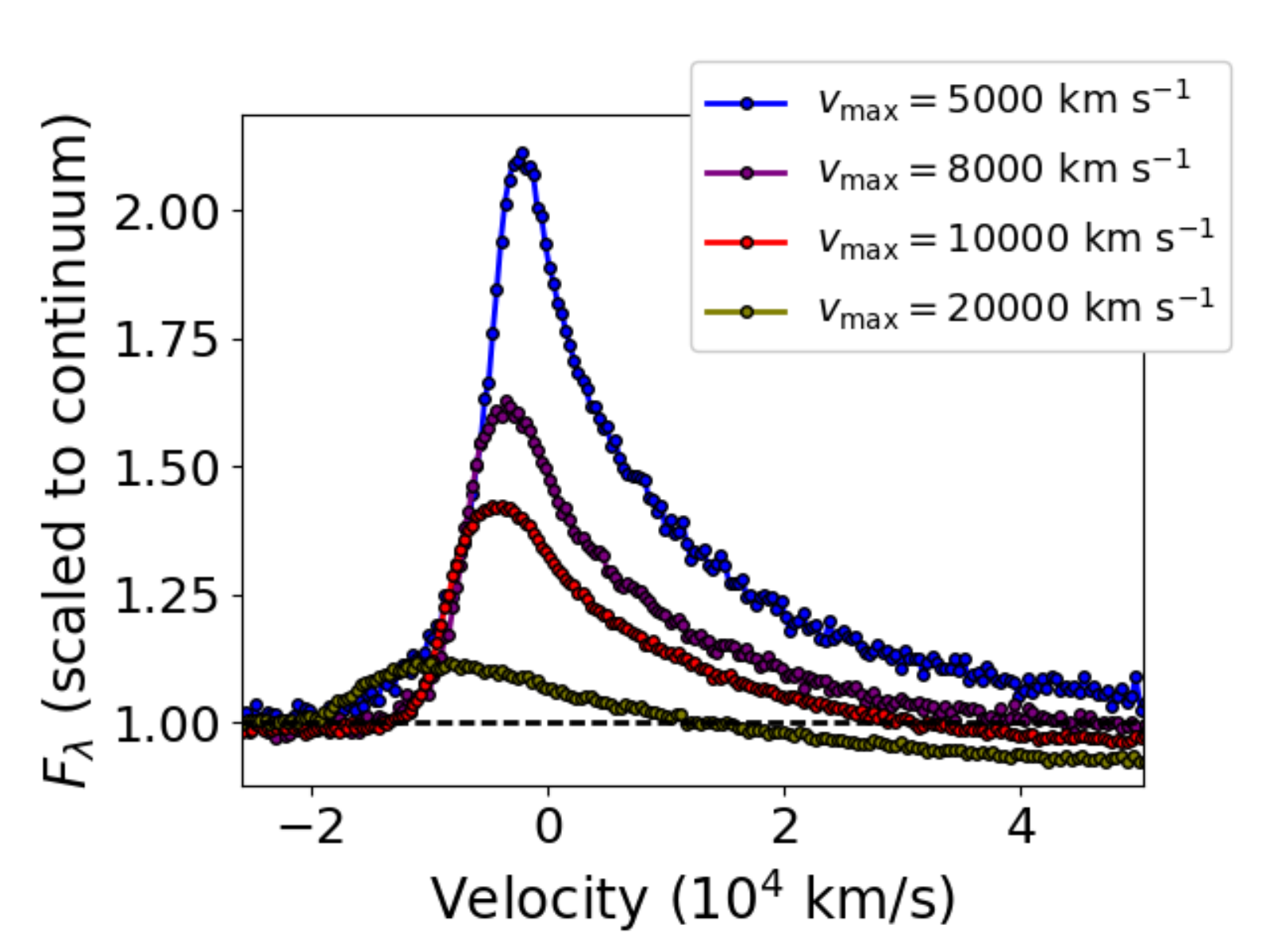}
\caption{Computed H$\alpha$ line profiles for a homologously expanding outflow  with various values of $v_{\rm sc}$. The continuum fluxes are, from lowest to highest $v_{\rm sc}$: $1.3$, $2.0$, $2.6$, and $6.5 \times 10^{38}$~erg~s$^{-1}$~$\AA^{-1}$. For all lines in this figure, the continuum photosphere $r_{\rm th}$ is located at $2.7 \times 10^{14}$~cm, and the outer radius $R = 10^{15}$~cm. }
\label{fig:VaryVmax}
\end{figure}

Figure~\ref{fig:VaryVmax} shows our more detailed  line profile calculations for  homologous outflow models with various values of $v_{\rm sc}$. The gas density and extent are set using the fiducial values introduced in Section~\ref{sec:TexVariation}. We set $T_{\rm in}$, the gas temperature at $r_{\rm in}$, equal to $2.93 \times 10^5$ K, chosen so that the diffusive luminosity of the fiducial envelope with $v_{\rm sc}= 10^{4}$ km s$^{-1}$ is $10^{45}$ erg s$^{-1}$. The continuum thermalization depth resides at $r_{\rm th} = 2.7 \times 10^{14}$~cm.

The first thing to note is that the lines are primarily in emission. There is no blueshifted absorption trough, as seen in the P-Cygni profiles associated with homologous outflow in a supernova. As explained in Section~\ref{sec:TexVariation} this is due to the high line source function found for lines in TDE outflows. The peak of the
model line profile is also  blueshifted, with a higher value of $v_{\rm sc}$ producing a larger blueshift. The lines are also asymmetric, with an extended red wing. These asymmetric lines profiles are similar to those studied in \citet{Auer1972}, \citet{Fransson1989}, and \citet{Hillier1991} .

Before proceeding to show more results, we will explain what causes these line shapes. We have already seen that, for a sufficiently high line excitation temperature $T_{\rm ex}$, lines that form in an expanding atmosphere appear purely in emission. We also saw that, in the absence of electron scattering, the lines possess a shoulder at Doppler velocity of the continuum photosphere, $v_{\rm th}$. The inclusion of electron scattering smooths the shoulder into a blueshifted peak.  Finally, the red tail of line emission arises because the photons are scattering in an expanding flow. Just as the continuum radiation is redshifted adiabatically in an outflow, a similar effect is seen on the line photons. 

\begin{figure}[htb!]
\includegraphics[width =0.5\textwidth]{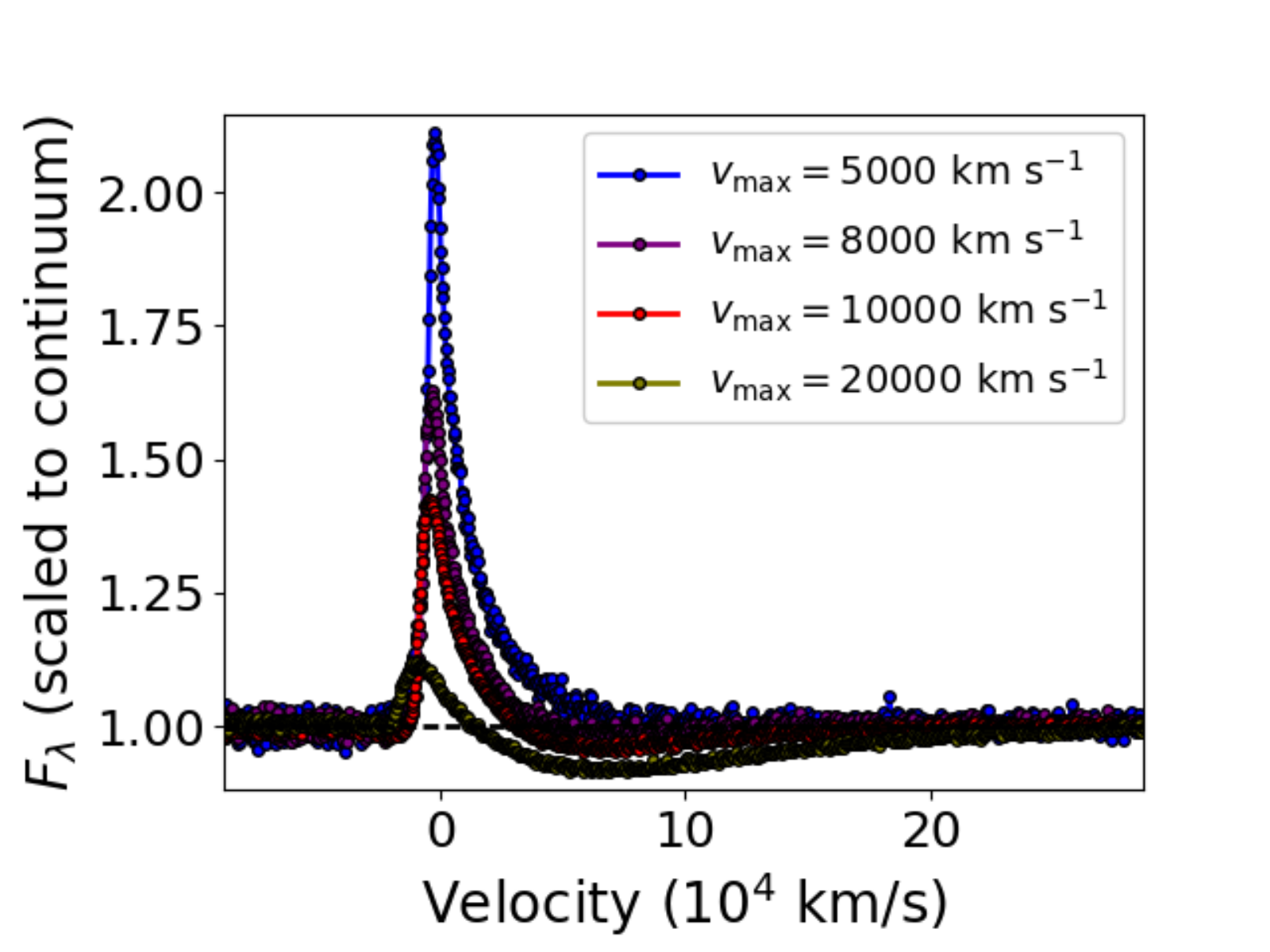}
\caption{The same as Figure~\ref{fig:VaryVmax}, but displayed over a wider range of wavelengths (Doppler velocities). At higher $v_{\rm sc}$, a redshifted absorption trough appears.}
\label{fig:VaryVmaxZoomedOut}
\end{figure}

Figure~\ref{fig:VaryVmaxZoomedOut} shows the same model spectra as Figure~\ref{fig:VaryVmax}, but zoomed out over a larger range of wavelengths. At higher $v_{\rm sc}$, a \emph{redshifted} absorption trough is visible. In our setup, however, it is not very prominent for $v_{\rm sc}$ less than $10^4$ km s$^{-1}$ and would be difficult to detect, given the signal-to-noise limitations in most spectra.

These line profiles bear a resemblance to the inverse P-Cygni profiles that result from the so-called ``top-lighting'' effect from ISM interaction in a supernova \citep{Branch2000}, but they arise here for different reasons. In the case of \citet{Branch2000}, the non-shell emission at each wavelength arises from constant projected velocity surfaces, which is not the case here because of the high $\tau_e$. In our case, the redshifted absorption is related to the overall adiabatic evolution of the continuum radiation. Starting at the inner boundary and moving out, the entire continuum is being redshifted as photons do work on the expanding envelope. When continuum radiation is absorbed by the line, the adiabatic redshifting transfers the absorption feature to longer wavelengths.

\subsection{Homologous Expansion with Different Outer Radii}

\begin{figure}[htb!]
\includegraphics[width =0.5\textwidth]{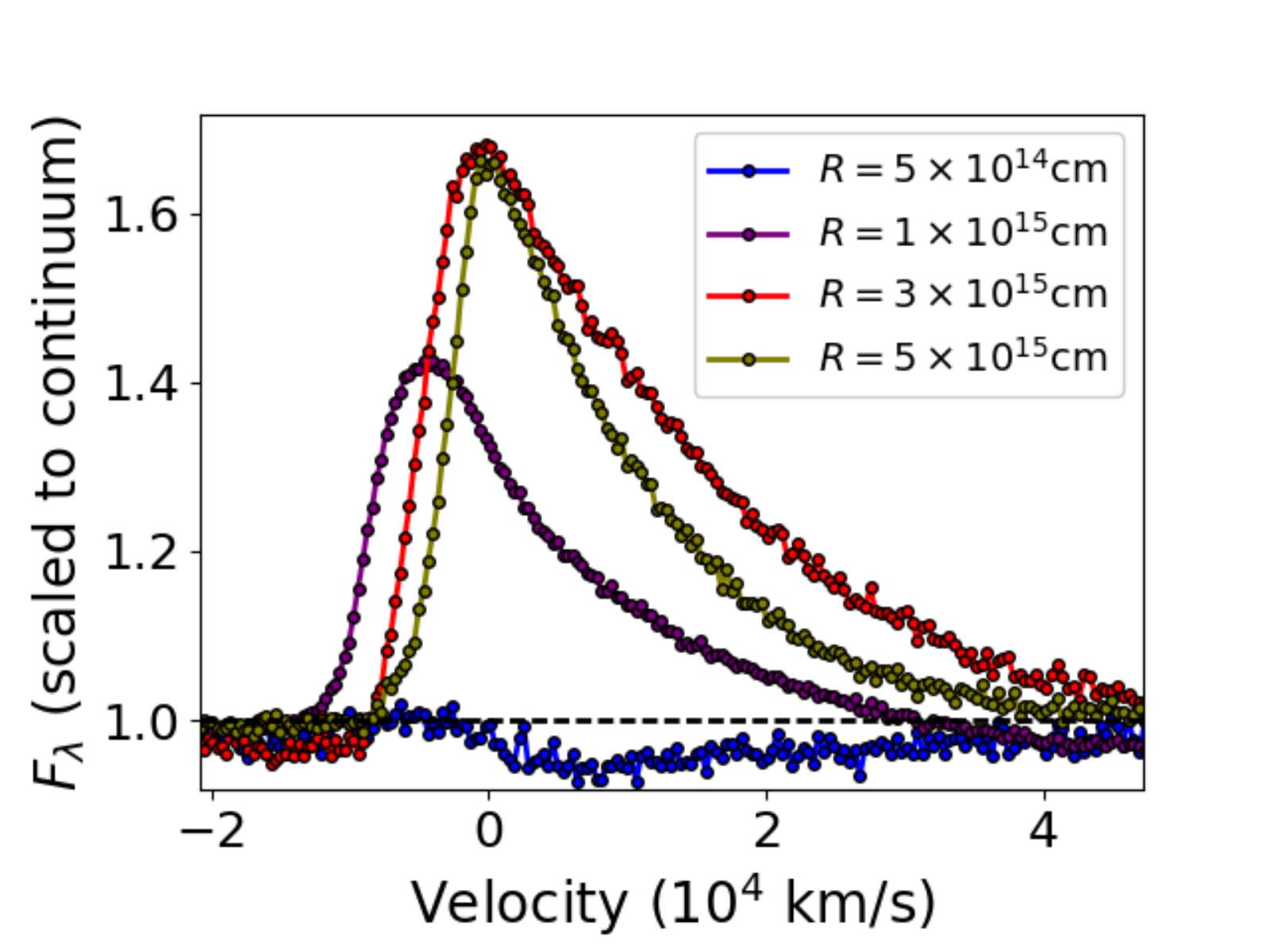}
\caption{Computed H$\alpha$ line profiles for homologous expansion and $ v_{\rm sc} = 10^4$ km s$^{-1}$, but for various values of the outer truncation radius $R$. This crudely represents a time sequence for a homologous outflow. The continuum fluxes are, from lowest to highest $R$: $7.3$, $2.6$, $0.91$, and $0.85 \times 10^{38}$~erg~s$^{-1}$~$\AA^{-1}$. The continuum thermalization photosphere $r_{\rm th}$ is located at 3.6, 2.7, 1.0, and 1.0 $\times 10^{14}$~cm, following the procedure described in Appendix~\ref{sec:RadiativeTransferSetup}.}
\label{fig:VaryRout}
\end{figure}

Figure~\ref{fig:VaryRout} shows the effect of varying $R$, while keeping $v_{\rm sc}$ fixed at $10^4$ km s$^{-1}$. We keep the envelope mass fixed at 0.25 $M_{\odot}$ and the diffusive luminosity fixed at $10^{45}$ erg s$^{-1}$ (the true bolometric luminosity will be affected by how the advective properties of the envelope change as we adjust its size). This set of calculations can be considered a crude representation of the time evolution of a TDE outflow where the radial extent of the outflow increases with time. We caution, however, that a time-dependent radiation-hydrodynamic calculation is necessary to truly model the time evolution.

For the smallest value of $R$ considered, $5 \times 10^{14}$~cm, the line profile becomes a shallow absorption, nearly blending into the continuum entirely. A similar effect was seen in R16 for a static envelope of otherwise similar parameters. At larger $R$, the emission reappears. As $R$ increases, the peak of the line becomes more centered, and at $R = 5 \times 10^{15}$~cm, the line is entirely centered on the rest wavelength. The line also narrows and becomes more symmetric as $R$ increases. 

\subsection{Constant-velocity versus Homologous Expansion}

\begin{figure}[htb!]
\includegraphics[width =0.5\textwidth]{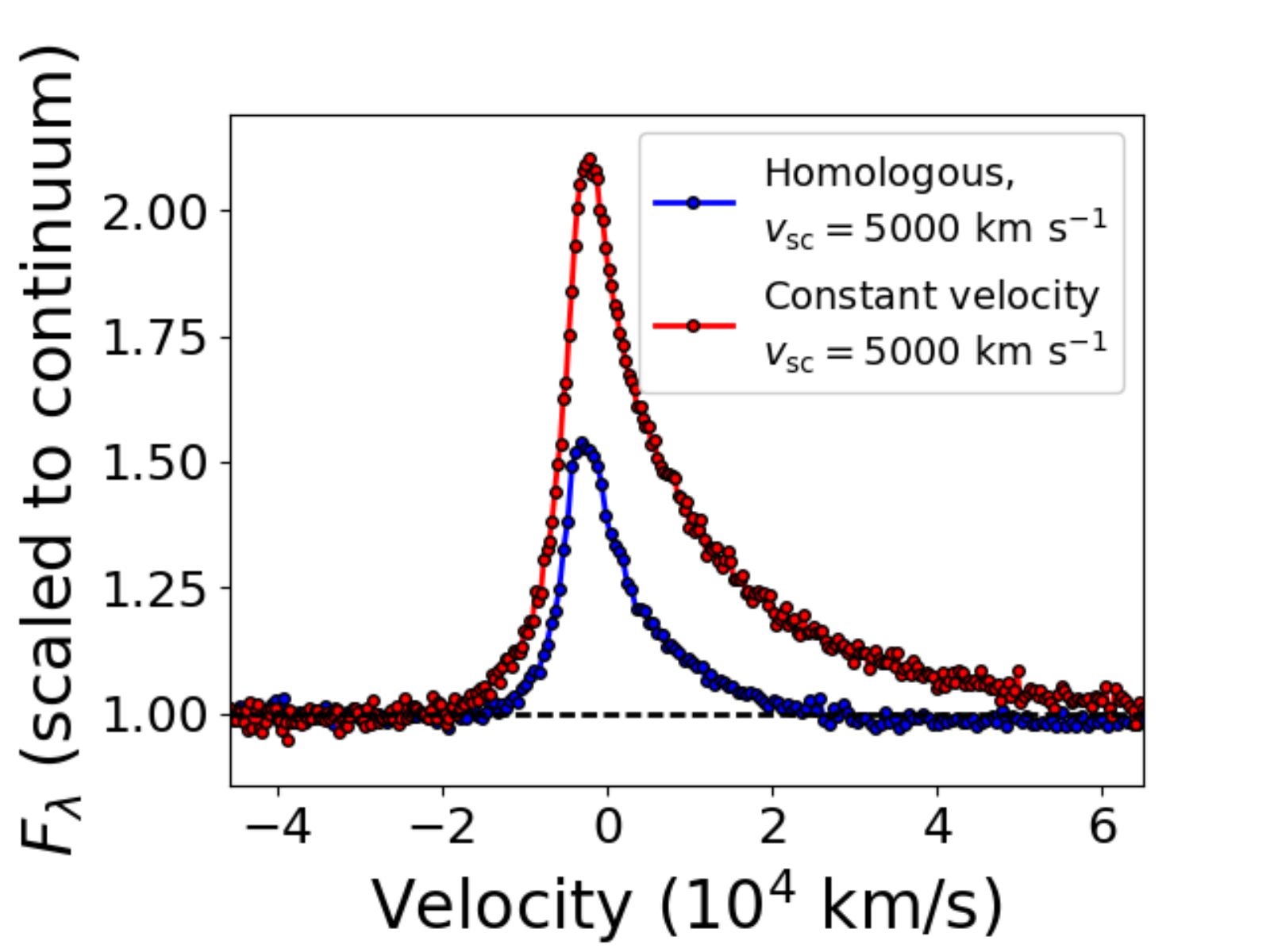}
\caption{Computed H$\alpha$ line profiles with $ v_{\rm sc} = 10^4$~km~s$^{-1}$, comparing the case of homologous expansion to the constant-velocity case. The continuum flux is $1.3 \times 10^{38}$~erg~s$^{-1}$~$\AA^{-1}$ for the homologous case, and $4.3 \times 10^{38}$~erg~s$^{-1}$~$\AA^{-1}$ for the constant-velocity calculation. The continuum thermalization photosphere $r_{\rm th}$ is $2.7\times 10^{14}$~cm for the homologous case and $2.9\times 10^{14}$~cm for the constant-velocity case. }
\label{fig:HomologousVsConstant}
\end{figure}

Figure~\ref{fig:HomologousVsConstant} compares the line profile of a homologous expanding model with $v_{\rm sc} = 10^4$~km~s$^{-1}$ to a model where the entire outflow moves with the same velocity $v_{\rm sc}$.  The line profiles are similar, but due to the enhanced adiabatic reprocessing in the constant-velocity case (see Appendix~\ref{sec:AdiabaticLossesWithEngine} for more details), the strength of the continuum is higher at the H$\alpha$ wavelength, reducing the contrast of the line in that case. 

\subsection{Comparison to ASASSN-14ae}
\label{sec:14aeComparison}

\begin{figure}[htb!]
\includegraphics[width =0.5\textwidth]{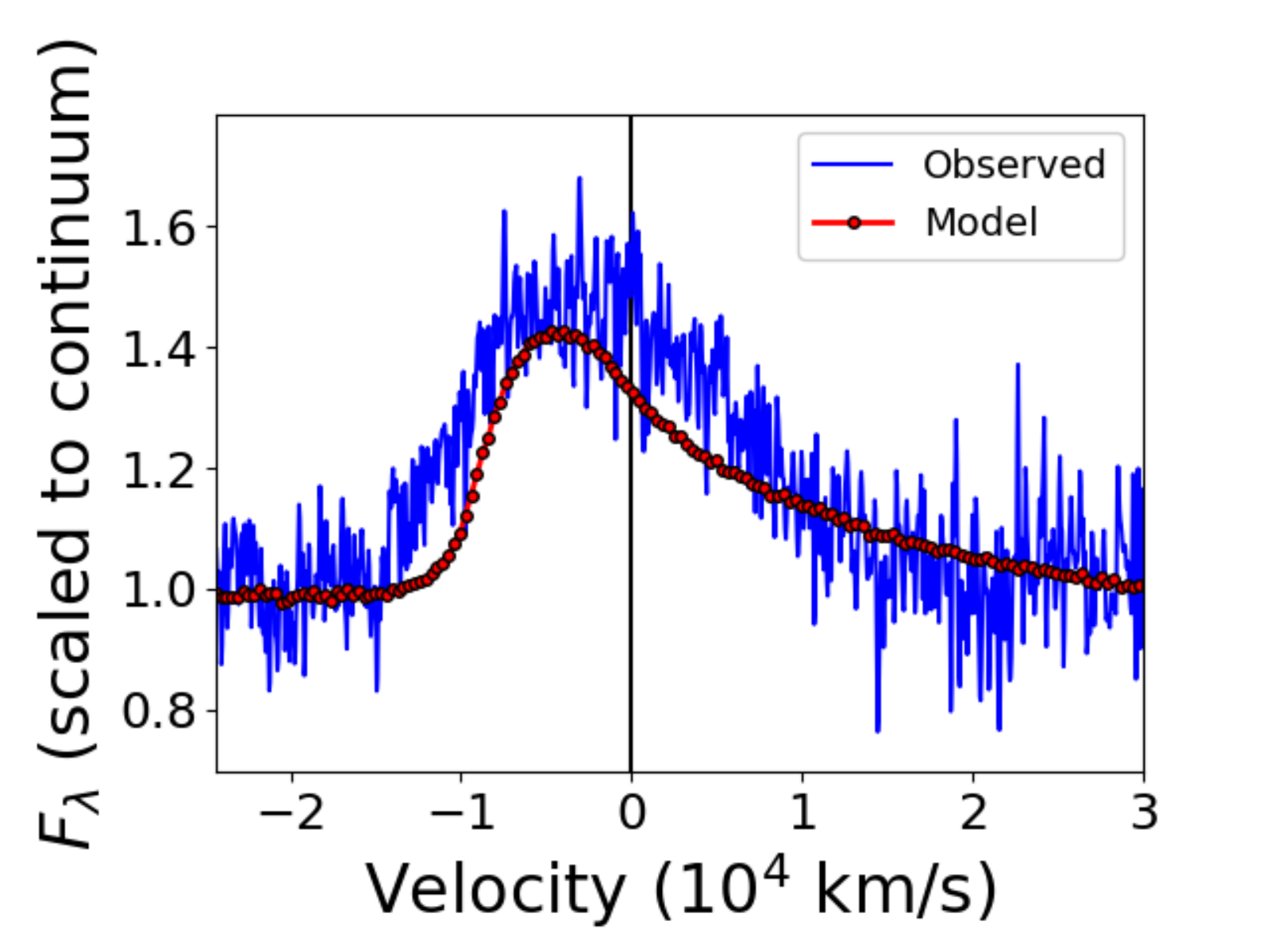}
\caption{
Comparison of the  observed H$\alpha$ line profile of ASASSN-14ae (blue curve, taken four days post-discovery) to a radiative transfer model of a homologously
outflow with $v_{\rm sc} = 10^4$ km s$^{-1}$ (red curve).  The model reproduces the key features of the line profile, including the 
blueshifted line peak and asymmetric red wing.   
The model continuum flux, however, is roughly a factor of 4.5 below the observed host-subtracted value at this epoch.} 
\label{fig:14aeComparisonEarly}
\end{figure}

Figure~\ref{fig:14aeComparisonEarly} displays how our fiducial H$\alpha$ line profile compares to the early (four days post-discovery) line profile observed from ASASSN-14ae \citep{Holoien2014}, accessed via the Weizmann interactive supernova data repository \citep{Yaron2012}. We succeed in obtaining a good match of the ratio of the peak line flux to that of the continuum. We also match a number of qualitative features of the line: its blueshifted peak, overall width, and asymmetry in the form of an extended red wing. The match is not perfect, however. The asymmetry in our computed line is more pronounced than in the observed one. Our line also does not extend as far to the blue as the observed one does. If we were to increase $v_{\rm sc}$, we would generate flux at bluer wavelengths, but at the cost of worsening all other aspects of the fit.

There is another important way in which our model falls short. While we match the \emph{relative} strengths of the line and the continuum, the value of our continuum flux at line-center, $2.6 \times 10^{38}$~erg~s$^{-1}$~$\AA^{-1}$, is about a factor of 4.5 too low compared to the observed host-subtracted value. The model was also designed to generate a bolometric luminosity of $10^{45}$ erg s$^{-1}$, whereas the estimated bolometric luminosity at this epoch, as reported in \citet{Holoien2014}, is about an order of magnitude lower. A more thorough exploration of parameter space might produce a fit that succeeds better in matching these aspects of the observations. In particular, the observation suggests a higher envelope mass and/or absorption reprocessing efficiency than we have assumed here. 

\section{Discussion and Conclusions}
\label{Sec:Discussion}

Asymmetric emission lines with blueshifted peaks and extended red wings can be signatures of outflows in TDEs. The H and He lines will generally not possess blueshifted absorption in the manner of a P-Cygni profile, because of the high excitation temperature of the lines. The red wing of the line is a result of the redshifting of line photons as they scatter through an expanding atmosphere. For a prompt outflow that expands with time, the initially blueshifted peak will become more centered, and the line asymmetry will decrease over time. These effects might help to explain the behavior of H$\alpha$ in ASASSN-14ae, as well as the behavior of the \ion{He}{2} line in PTF-09ge, ASASSN-15oi, and PS1-10jh, all of which display emission that is more blueshifted than expected for the Bowen blend near this line. The red wing produced by this mechanism might also help to explain the asymmetric H and \ion{Fe}{2} line profiles seen in the TDE candidate PS1-16dtm \citep{Blanchard2017}.

Electron scattering can play a significant role in setting the width of the emission lines: in the absence of an outflow or un-attenuated emission from an accretion disk, it may be the dominant source of the width. The narrowing of emission lines over time, as has been observed in events including ASASSN-14li and ASASSN-14ae, might be more easily explained in terms of an evolution in the optical depth of the line-emitting region over time, rather than an evolution in its kinematics. We have tested this idea by performing fits to the ASASSN-14li H$\alpha$ line profiles, varying only $\tau_e$, and obtaining fits at three epochs. We remain agnostic at this time as to what hydrodynamic processes may cause the optical depth to drop over time. The role of electron scattering in shaping spectral features could be further tested via spectropolarimetry, as suggested by \citet{Chugai2001}.

Our interpretation of the line profiles in ASASSN-14li is complicated by the fact that radio observations of this event suggest that it led to the launch of a wide-angle outflow \citep{Alexander2016}. Given our results for line profiles in moving atmospheres, we might therefore expect the lines in ASASSN-14li to display the asymmetries we studied in Section~\ref{sec:CombinedCalculations}. These findings may be reconciled if the geometry of the emitting material is non-spherical, such that we are seeing line emission along a line of sight that intercepts non-outflowing material. The simultaneous x-ray emission from this event also hints at the presence of multiple emitting surfaces, although it is not clear whether the fact that we see the x-rays is consistent with the suggestion that the outflow is hidden from view. Meanwhile, the radio data from 14li has also been interpreted as resulting from a narrow jet \citep{van-Velzen2016-1}, or from the unbound stellar debris of the star \citep{van-Velzen2016-1}, which in both cases could be consistent with the conclusion that most of the line-emitting gas is not outflowing.

In the presence of an outflow, a sufficiently compact reprocessing envelope can still lead to the near-total suppression of the H$\alpha$ line with respect to the continuum, similar to what was found for a static envelope in R16, and relevant to TDEs such as PS1-10jh which show no detectable hydrogen emission in their spectra. However, we do see some evidence that, as the outflow proceeds and the envelope expands, the strength of the hydrogen line with respect to the continuum may change.

Though the models in this paper help illuminate several key features of line formation in TDE outflows, we have made a number of assumptions and simplifications that will need to be improved upon in future work. We have assumed spherical symmetry, which prevents us from accounting for viewing angle effects. While we have accounted for radial motion of the gas in an outflow, we have not included rotational motion, which in some scenarios may be of comparable magnitude. Our treatment is time-independent ,in the sense that we assume the radiation diffusion time is small compared to the hydrodynamic timescales, which may not be true---especially at times before the light curve peak. 

To determine the gas density and velocity as a function of position and time, rather than treating these quantities parametrically as we have done here, we would need to perform radiation-hydrodynamics simulations in three spatial dimensions.

Despite these shortcomings,  the trends we have described here are likely to be qualitatively robust and to pave the way toward a more complete understanding of the optical and UV emission from TDEs. 

\section*{Acknowledgments}

We thank Iair Arcavi, Nadia Blagorodnova, Jon S. Brown, Brad Cenko, Suvi Gezari, Tom Holoien, Julian Krolik, and Brian Metzger for helpful conversations. N.R. acknowledges the support of a Joint Space-Science Institute prize postdoctoral fellowship. 
D.K. is supported in part by a Department of Energy Office of Nuclear
Physics Early Career Award, and by the Director of the Office of Energy
Research, Office of High Energy and Nuclear Physics, Divisions of
Nuclear Physics, of the U.S. Department of Energy, under Contract No.
DE-AC02-05CH11231. 
This work was supported in part by NSF Astronomy and Astrophysics grant 1616754.
Simulations were performed on the Deepthought2 high-performance computing cluster at the University of Maryland.

\appendix

\section{Treatment of the Continuum Radiation and Line Source Functions}
\label{sec:Method}

\subsection{Scope}
\label{sec:Scope}
In principle, as in R16, we need to simultaneously (iteratively) solve the non-LTE equations coupled with the radiative transfer equation. The former determine the emissivity and opacity of each line, while the latter determines quantities such as photoionization rates and mean radiative intensities at line wavelengths $\bar{J}_{\lambda,\rm line}$, which enter into the non-LTE equations that determine the line emissivities and opacities.

However, for computational expediency, in this study we forgo the iterative approach. We instead solve the non-LTE equations while assuming that the continuum radiation is entirely responsible for setting the line emissivities and opacities. In other words, we assume that $\bar{J}_{\lambda,\rm line}$ is equal to the value of $J_\lambda$ at the neighboring continuum. Given that observations indicate that the flux at line center in TDEs is generally within a factor of a few of the neighboring continuum flux (with the notable exception of ASASSN-14li), we feel that this is a reasonable approximation. To the extent that this approximation fails, as it is increasingly likely to do when applied to spectra taken at later times, it will introduce quantitative errors into our predictions for the line ratios and widths, but we should still be able to discern qualitative patterns.

We also simplify our calculation of the properties of the continuum radiation. We track two effects: (1) Adiabatic reddening of the spectrum injected at the lower boundary of the envelope, and (2) Absorption of soft x-ray and UV photons, followed by emission at longer wavelengths. We describe our treatment of the first effect in Appendix~\ref{sec:AdiabaticLossesWithEngine}, and of the second in Appendix~\ref{sec:TwoTemperatureReprocessing}. We then go on to describe how we can translate these properties of the continuum radiation into line opacities and source functions in Appendix~\ref{sec:NLTE}. We provide additional details for how we use this information in our radiative transfer calculations in Appendix~\ref{sec:RadiativeTransferSetup}. 

\subsection{Radiation Energy Density as a Function of Radius: Role of Central Engine and Adiabatic Losses}
\label{sec:AdiabaticLossesWithEngine}

We consider a fluid in which radiation dominates its internal energy. We assume that the gas is dense enough that the radiation is in the diffusion regime (but not necessarily in local thermodynamic equilibrium). To order $v/c$ (where $v$ is the fluid velocity and $c$ is the speed of light), and in spherical symmetry with radial coordinate $r$, the first law of thermodynamics for a fluid element is expressed by \citep{Mihalas1984}
\begin{equation}
  \label{eq:LagrangianEnergy}
   \frac{D}{Dt} \left(\frac{E_0}{\rho}\right) + \frac{1}{3}E_0 \frac{D}{Dt}\left(\frac{1}{\rho}\right) = - \frac{1}{4 \pi r^2 \rho}\frac{\partial L_0}{\partial r} \, \, ,
\end{equation}
where $E_0$ is the radiation energy density per unit volume, as measured in the co-moving frame of the fluid; $\rho$ is the fluid mass density, $L_0 = 4 \pi r^2 F_0$, $F_0$ is the co-moving radiative flux; and $D/Dt$ is the Lagrangian (material) derivative operator. We have taken the radiation pressure in the diffusion regime to be equal to $1/3 E_0$.

We consider a medium that is ionized highly enough that electron scattering dominates the opacity, which we denote by $\kappa_{\rm es}$, with corresponding optical depth $\tau_e$. This opacity is evaluated in the co-moving frame of the fluid. The condition of radiative diffusion then allows us to write
\begin{equation}
  \label{eq:RadiativeDiffusion}
  \frac{L_0}{4 \pi r^2} = -\frac{1}{3} \frac{c}{\rho \kappa_{\rm es}} \frac{\partial E_0}{\partial r} \, \, .
\end{equation}
We also make use of the continuity equation 
\begin{equation}
  \label{eq:ContinuityLagrangian}
  \frac{D \rho }{Dt} = -\rho \left(\vec{\nabla} \cdot \vec{v}\right) \, \, .
\end{equation}
Combining all of these and using spherical symmetry to expand the $\nabla \cdot \vec{v}$ terms gives 
\begin{equation}
  \label{eq:CombinedEquation}
\frac{\partial{E_0}}{\partial t} + v\frac{\partial{E_0}}{\partial r}  + \frac{4}{3} E_0 \left(\frac{\partial{v}}{\partial r} + \frac{2 v}{r} \right) = \frac{1}{r^2}\frac{\partial}{\partial r} \left( \frac{c}{3 \rho \kappa_{\rm es}} r^2 \frac{\partial E_0}{\partial r} \right) \, \, .
\end{equation}

At this point, we will drop the $0$ subscript for the radiation energy density, with the understanding that $E$ always refers to the co-moving radiative energy density. To convert to the lab-frame value of $E$, we can use the relation 
\begin{eqnarray}
  \label{eq:LabFrameEnergyDensity}
E_{\rm lab} &=& E_0 + 2 \frac{v}{c} \frac{F_0}{c} \nonumber \\
&=& E_0 -  \frac{2v}{3 \rho \kappa_{\rm es} c} \frac{\partial E_0}{\partial r} \, \, ,
\end{eqnarray}
which is accurate to order $v/c$ \citep{Mihalas1984}. The second equality makes use of the diffusion approximation.

Next, we follow \citet{Arnett1980} (hereafter A80) by assuming the solution for $E$ is separable in space and time, and factoring out the adiabatic dependence on $R(t)$,
\begin{equation}
  \label{eq:SeparateE}
  E(r,t) = E(r_{\rm in},0)\psi(x)\phi(t)\frac{R^4(0)}{R^4(t)} \, \, ,
\end{equation}
where $x \equiv r / R(t)$. We obtain
\begin{equation}
  \label{eq:PreSeparationEquation}
  R(t)\frac{\dot{\phi}}{\phi} - 4 \dot{R} + v\frac{\psi^\prime}{\psi} + \frac{4}{3}\left(v^\prime + \frac{2v}{x}\right) = \frac{c}{3 R(t) }\frac{1}{\psi x^2} \frac{\partial}{\partial x}\left(\frac{x^2}{\rho \kappa_{\rm es}}\psi^\prime\right) \, \, ,
\end{equation}
where dots denote partial derivatives with respect to time, and primes denote partial derivatives with respect to $x$. 

If we were to continue following A80, we would assume homologous expansion in the form $v = v_{\rm sc} x$, and  for an appropriate scale velocity $v_{\rm sc}$. This leads to a cancellation of the second and fourth terms of equation~\eqref{eq:PreSeparationEquation}, resulting in  
\begin{equation}
  \label{eq:HomologousCombined}
 R(t)\frac{\dot{\phi}}{\phi} + v_{\rm sc} x\frac{\psi^\prime}{\psi} = \frac{c}{3 R(t) }\frac{1}{\psi x^2} \frac{\partial}{\partial x}\left(\frac{x^2}{\rho \kappa_{\rm es}}\psi^\prime\right) \, \, .
\end{equation}
The second term of equation~\eqref{eq:HomologousCombined} does not appear in A80. In our application, this term will be important, so we proceed differently. If the radiation diffusion time is small compared to the time over which the envelope properties change, then the terms containing partial time derivatives in equation~\eqref{eq:PreSeparationEquation} are small compared to the other terms. Dropping those terms, and only those terms, we are left with
\begin{equation}
  \label{eq:SpatialPartOnly}
  v \psi^\prime + \frac{4}{3} \left(v^\prime + \frac{2v}{x} \right) \psi =\frac{c}{3 R(t)} \frac{1}{x^2}\frac{\partial}{\partial x} \left( \frac{x^2}{\rho \kappa_{\rm es}} \psi^\prime \right)  \, \, .
\end{equation}

To proceed further, we need $v(r)$ and $\rho(r)$. These are set by the hydrodynamics, particularly through the inclusion of the momentum conservation equation; in principle we need to solve for them simultaneously. Such solutions (for the time-independent case) have been described in \citet{Shen2016}. Alternatively, we can specify guesses for these in advance. For example, we can again consider homologous expansion, so that $v = v_{\rm sc} r / R$.  We can also consider a constant-velocity case where $v = v_{\rm sc}$ at all radii in our computational domain (i.e. the gas was initially accelerated at unresolved radii). These velocity profiles are within the range of outcomes of the \citet{Shen2016} solutions, in which $v \propto r$ at small radii and asymptotes to a constant at large radii. We will assume that $\rho$ can be written as a generic function of $x$. We introduce one final non-dimensional variable $\eta \equiv \rho(x) / \rho_{\rm in}$ where the subscript ``in'' refers to the value at the inner boundary. We obtain
\begin{subequations}
  \label{eq:NondimensionalWithVelocityProfile}
\begin{align}
 \psi^{\prime \prime} &= \left(\frac{\eta^\prime}{\eta} - \frac{2}{x} \right) \psi^{\prime} + \alpha \eta \left(x \psi^\prime + 4 \psi \right) \qquad \; \; \, {\rm (homologous)}\\
 \psi^{\prime \prime} &= \left(\frac{\eta^\prime}{\eta} - \frac{2}{x} \right) \psi^{\prime} + \alpha \eta \left(\psi^\prime + \frac{8}{3} \frac{\psi}{x} \right) \qquad {\rm (constant \, \, velocity)} \, \, ,
\end{align}
\end{subequations} 
where 
\begin{equation}
\label{eq:AlphaDefinition}
\alpha \equiv 3 \frac{v_{\rm sc}}{c} \rho_{\rm in} \kappa_{\rm es} R \, \, .
\end{equation}
We see here that $\alpha$ encodes information about both the optical depth of the envelope and the gas dynamical time. 

We treat the density as a power law, $\eta = x^{-n} (R / r_{\rm in})^{-n}$, that is truncated at radius $R$. The density power law $n$ and the value of $\rho_{\rm in}$ are free parameters. 

Now we must specify boundary conditions. As in A80, the Eddington approximation for the outer boundary results in 
\begin{equation}
  \label{eq:OuterBoundary}
\psi(1) = -\frac{2}{3} \psi^{\prime}(1)\frac{1}{\rho(R) \kappa_{\rm es} R} \, \, .
\end{equation}
Finally, we need to specify the flux emanating from the inner boundary by finding the appropriate value for $\psi^\prime$ at the inner boundary. In the supernova situation, that flux is usually taken to be zero, but here we are expecting a large luminosity to coming from the TDE engine. From the diffusion equation for the radiation energy density (equation~\eqref{eq:RadiativeDiffusion}), we have 
\begin{equation}
  \label{eq:InnerFlux}
 \psi^{\prime}(x_{\rm in}) \approx - \frac{3 \rho \, \kappa_{\rm es} \, L}{4 \pi \, c \, r_{\rm in}^2} \, \, .
\end{equation}
However, to find the exact value of the inner flux, we must solve the two-point boundary value problem. We proceed via the shooting technique. We start at the outer boundary, with a guess for $\psi(1)$. We use the outer boundary condition~\eqref{eq:OuterBoundary} to find $\psi^\prime$ there. Next, we solve equation~\eqref{eq:NondimensionalWithVelocityProfile}, moving to smaller radii until we reach the inner boundary. By definition, 
\begin{equation}
  \label{eq:InnerBoundary}
  \psi(x_{\rm in}) = 1 \, \, ,
\end{equation}
 so we adjust our guess for $\psi(x_{\rm out})$ until this is achieved. In so doing, we find the appropriate value of not only $\psi(1)$, but also of $\psi^\prime$ at the inner boundary. 

While the solution described above sets the ratio between $\psi$ at the inner and outer boundaries, the physical value of the radiation energy density at the inner boundary, $E_{\rm in}$, remains a free parameter. The physical values for $r_{\rm in}$ and $R$ are also free parameters. At the inner boundary, the radiation spectrum is a blackbody with temperature $T_{\rm in}$, where $E_{\rm in} = a_{\rm rad} T_{\rm in}^4$.

\begin{figure}[htb!]
\plotone{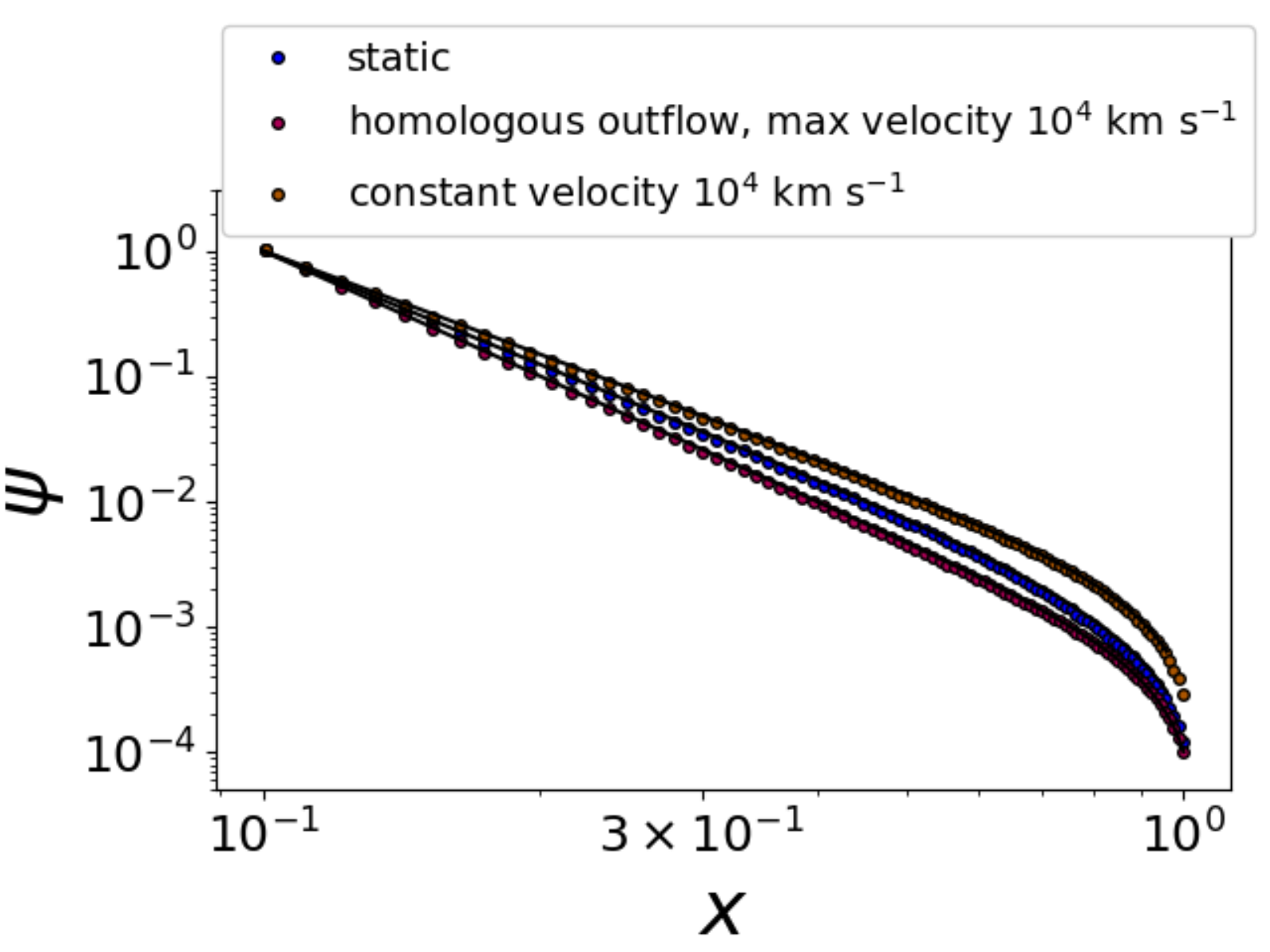}
\caption{Nondimensional radiation energy density $\psi$ as a function of nondimensional radius $x$. Circle markers are output from Monte Carlo radiative transfer calculation divided into 1024 radial zones, with data from every twelfth zone plotted. The solid black lines are solutions to the ordinary differential equation~\eqref{eq:NondimensionalWithVelocityProfile} subject to the boundary conditions from equations~\eqref{eq:OuterBoundary} and \eqref{eq:InnerBoundary}. These solutions assume that the radiation diffusion time is small compared to the time over which the envelope properties change. The maximum velocity for the homologously expanding envelope is $10^4$~km~s$^{-1}$, which is also the velocity used for the constant-velocity envelope, and both these envelopes have $\alpha = 152$.}
\label{fig:EnergyDensity}
\end{figure}

Figure~\ref{fig:EnergyDensity} shows solutions for $\psi(x)$ for three different guesses for the velocity structure. The other parameters, taken to be the same for all three curves, are $r_{\rm in} = 10^{14}$~cm, $R = 10^{15}$~cm, $v_{\rm sc} = 10^4$~km~s$^{-1}$, density power $n = 2$, and $\rho_{\rm in} = 4.32 \times 10^{-12}$~g~cm$^{-3}$. Combined with the aforementioned parameters, this implies an envelope mass of $0.25$~$M_\odot$. The electron scattering opacity $\kappa_{\rm es}$ is set for a fully ionized envelope consisting of hydrogen and helium at a solar abundance ratio, which evaluates to 0.34~cm$^2$~g$^{-1}$. This results in an electron scattering optical depth $\tau_e = 137$. The resulting value of $\alpha$ is 152 for the homologous and constant-velocity envelopes.

To better understand Figure~\ref{fig:EnergyDensity}, we can track how the components of the luminosity vary with radius in these models. Through each spherical shell of the envelope, there will be a flux of both advected radiative luminosity and diffusing radiative luminosity. There will also be a portion of the radiative energy that is lost to adiabatic work on the gas. The sum of these three components should be constant at each position. 

To see why, we can combine equations \eqref{eq:LagrangianEnergy} through \eqref{eq:ContinuityLagrangian} to obtain an energy conservation equation in conservative form:
\begin{equation}
  \label{eq:FirstLawConservationForm}
  \frac{\partial E}{\partial t} + \vec{\nabla} \cdot \left(\vec{v}E + \vec{F}_0  \right) = -p_{\rm rad} \vec{\nabla}\cdot \vec{v}
\end{equation}
From the divergence theorem, we see that the two fluxes through a shell boundary (the divergence term) are balanced by the volume integral of the radiation pressure term. Thus, we have

\begin{eqnarray}
L_{\rm diff} &=& -4 \pi r^2 \frac{c}{3 \rho \kappa_{\rm es}} \frac{dE}{dr} \nonumber \\
L_{\rm adv} &=& 4 \pi r^2 v E \nonumber \\
dW_{\rm adb}/dt &=& \int p_{\rm rad} \left(\vec{\nabla}\cdot \vec{v}\right) dV \nonumber \\
            &=& \int 4 \pi r^2 \frac{1}{3} E \left(\frac{\partial{v}}{\partial r} + \frac{2 v}{r}\right) dr \,\, .
\end{eqnarray}

\begin{figure}[htb!]
\plottwo{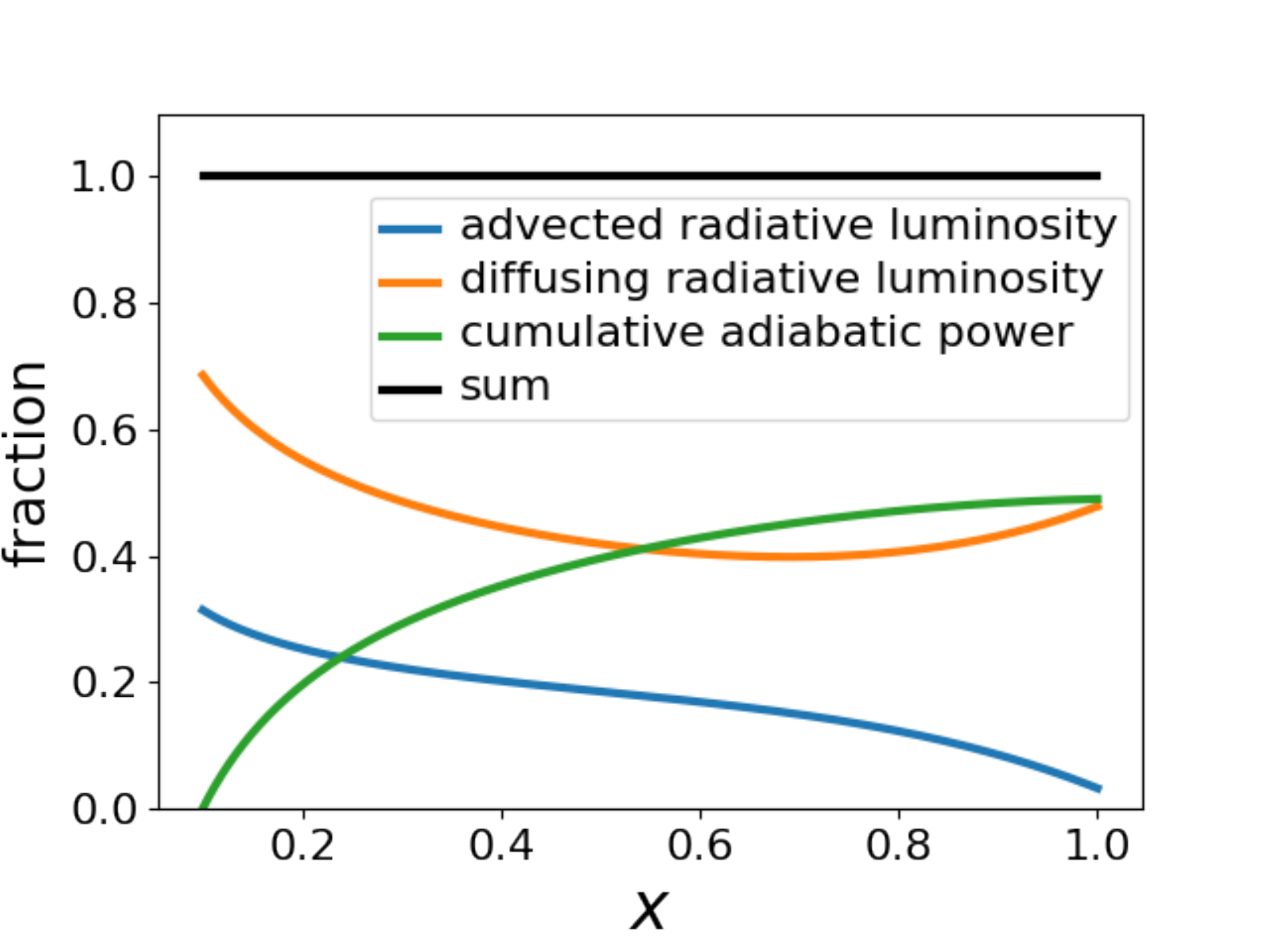}{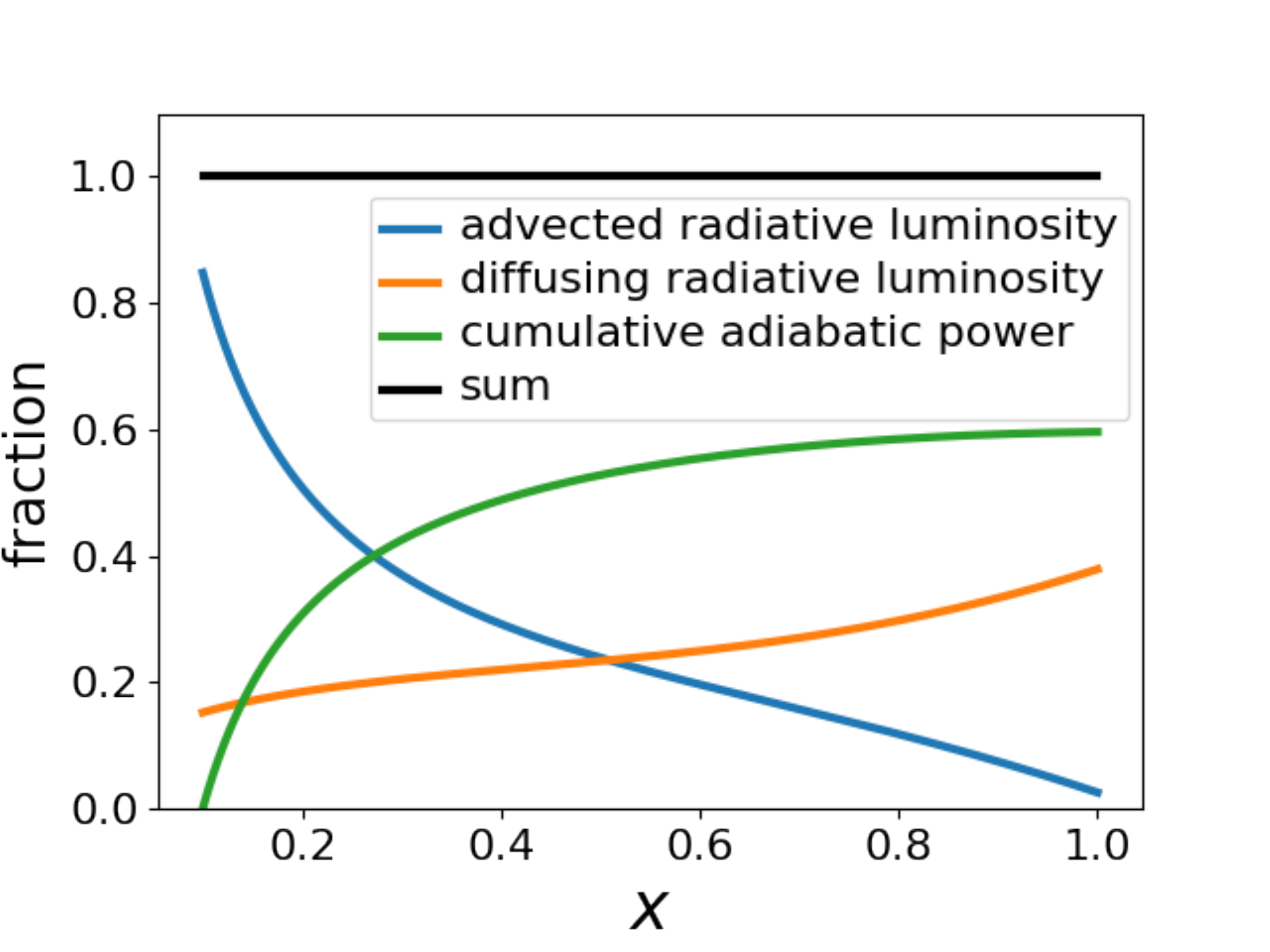}
\caption{The components of luminosity as a function of radius, for (left) a velocity profile corresponding to homologous expansion and (right) a constant-velocity outflow.}
\label{fig:EnergyConservation}
\end{figure}

Figure~\ref{fig:EnergyConservation} shows the luminosity components for the envelope described above undergoing homologous expansion as described above, and for a constant-velocity envelope. The analogous plot for the static envelope would show the orange curve overlapping with the black curve, with the green and blue curves at zero.

In both figures, nearly all of the radiative transfer at the outer boundary is diffusive. For the constant-velocity envelope, the energy transfer at small radii is primarily advective, while diffusion takes over at about $x = 0.5$. This corresponds to what is termed the ``trapping radius'' in the accretion literature \citep{Begelman1978,Meier1982-1}, and to the radiation breakout radius in the supernova literature \citep{Chevalier1992}, where the radiation diffusion time through the remaining envelope becomes comparable to the gas dynamical time $r/v$. For our setup, no trapping radius exists for the homologous case, as diffusion dominates the energy transfer at all radii.

The total radiative luminosity represented by the flat black curve does not correspond to the same value in the two figures. These calculations were set up to have the same radiative energy density at the inner boundary. The flux at the inner boundary adjusts as a result of the solution of the two-point boundary value problem, such that it is larger for the constant-velocity case than the homologous expansion case. This is a consequence of the large advective flux of radiative energy at the inner boundary of the constant-velocity calculation. This helps to explain why the radiative energy density in Figure~\ref{fig:EnergyDensity} is highest for the constant-velocity case, even when accounting for radiative energy loss to adiabatic expansion. Meanwhile, for the homologous case, where the advective flux at the inner boundary is much lower, the adiabatic losses result in a lower radiative energy density than for the static case.

\begin{figure}[htb!]
\plotone{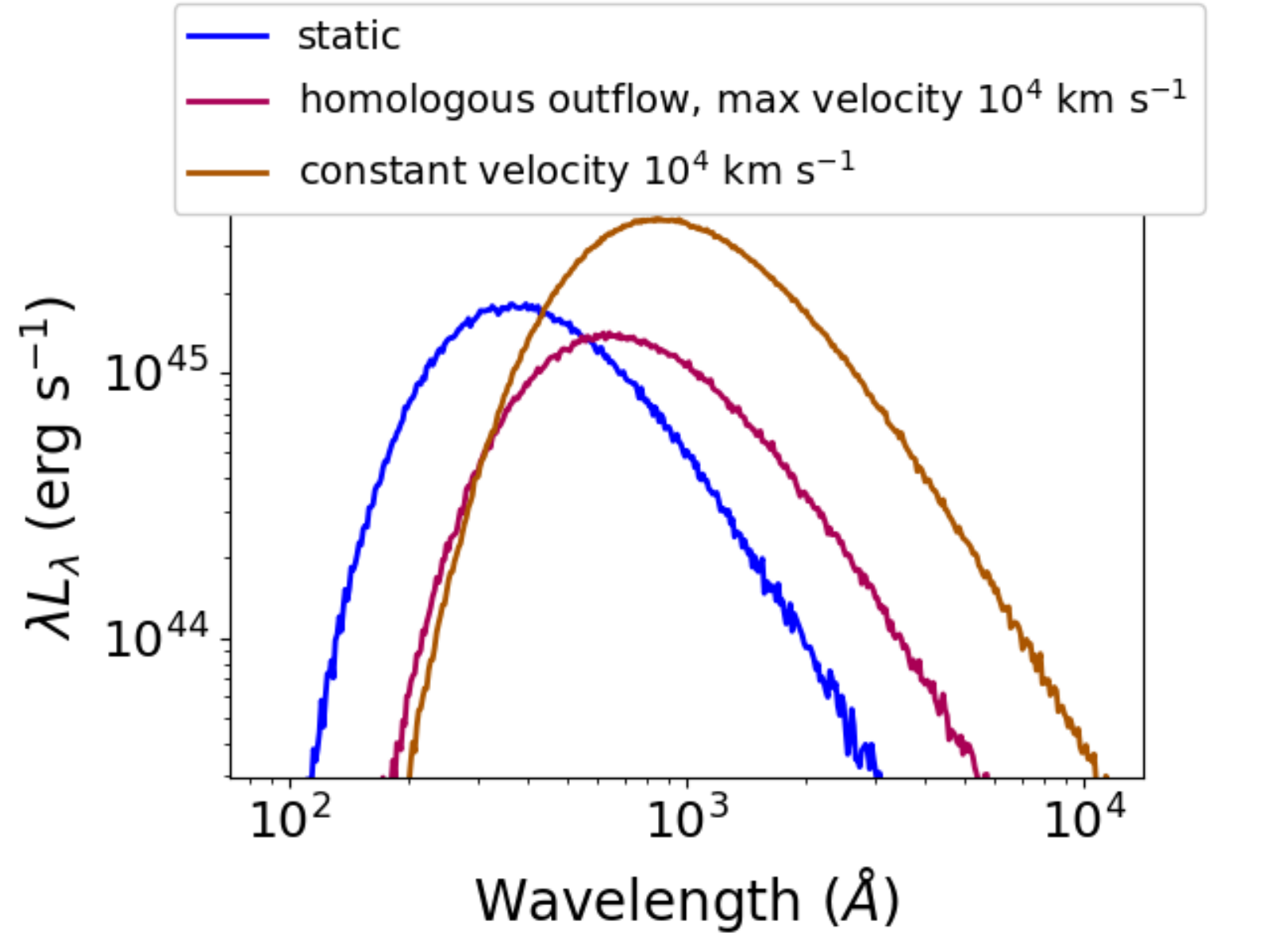}
\caption{The effect of adiabatic reprocessing on the escaping SED for the three velocity structures considered. In all three cases, a blackbody with temperature of approximately $2.93 \times 10^5$ K was injected at the inner boundary; for the given envelope density profile, this corresponds to a diffusive luminosity of $10^{45}$ erg s$^{-1}$ at the outer boundary for the homologous atmosphere. Expanding atmospheres lead to a redshift of the injected SED, and also influence the total luminosity that escapes. The radiation energy density at the inner boundary is set to be the same in all three cases, but this means that the inner luminosity in all three cases is different.}
\label{fig:SpectraRedshifting}
\end{figure}

We can use the information displayed in Figure~\ref{fig:EnergyConservation} to understand how the spectrum of the radiation evolves as a function of position. The energy lost to adiabatic expansion leads to a redshifting of the spectrum. We can see this in the emergent SEDs of the three models, displayed in Figure~\ref{fig:SpectraRedshifting}. The SED peaks at longer wavelengths for the homologous and constant-velocity calculations than for the static calculation. We see that adiabatic reprocessing can make a substantial contribution to the flux that escapes at optical wavelengths, as has been discussed in past work \citep{Strubbe2009,Lodato2011,Metzger2016}.

\subsection{A Simplified ``Two-temperature'' Reprocessing Scheme}
\label{sec:TwoTemperatureReprocessing}

In R16 we demonstrated how a TDE envelope can absorb soft x-ray and UV radiation emitted from accretion onto the black hole (BH) and reprocess them to longer wavelengths. The optical continuum is ultimately composed of a blend of emission from different temperatures originating from different radii within the envelope; its strength depends on details of the envelope structure such as its density and radial extent. Here, we will collapse all of these details into an approximation formula that depends on two parameters. The first parameter is $\epsilon_{\rm abs}$, which acts as an average ratio of absorption opacity to electron-scattering opacity for UV and soft x-ray photons. The second parameter, $f$, denotes the fractional temperature of the reprocessed radiation compared to $T_{\rm in}$.

Consider again the effects of adiabatic expansion, as described in the previous section. The spectrum at the inner boundary is a blackbody, $B(\lambda,T_{\rm in})$. At larger radii, the spectrum is given by
\begin{equation}
  \label{eq:RedshiftedSpectrum}
J_\lambda(r) = J(r) \, B\left[d(r)\lambda ,T_{\rm in}\right] \, \, ,
\end{equation}
where $d(r)$ is the photon degradation factor, defined as the mean photon energy at that radius divided by its energy at the inner boundary. In other words, $d(r) = 1 - w_{\rm ad}(r)$, where $w_{\rm ad}$ is the fraction of luminosity lost to adiabatic work, as displayed by the green curves in Figure~\ref{fig:EnergyConservation}. The normalization factor, $J(r)$, out front ensures that the wavelength-integrated radiation energy density as a function of radius matches what we found in Appendix~\ref{sec:AdiabaticLossesWithEngine}, as displayed in Figure~\ref{fig:EnergyDensity}. We now incorporate the effect of UV/x-ray absorption and re-emission at longer wavelengths, as follows:
\begin{eqnarray}
  \label{eq:TwoTemperatureReprocessing}
J_\lambda^\prime(r) &=& J(r) ( B[d(r) \lambda, T_{\rm in}] e^{-\sqrt{\epsilon_{\rm abs}} \tau_e} \nonumber \\
&+& f^{-4} B[d(r) \lambda, f T_{\rm in}](1 -  e^{-\sqrt{\epsilon_{\rm abs}} \tau_e}) ) \, \, .
\end{eqnarray}
For this single equation, $\tau_e$ is integrated from the inner boundary outward. For all the calculations in this paper, we will set $\epsilon_{\rm abs} = 10^{-5}$ and $f = 1/2$. These values are informed by the full non-LTE calculations of R16. In reality, these reprocessing parameters will depend on the other parameters such, as $R$, $\rho_{\rm in}$, etc., but we choose not to account for this added complication at this time.

\begin{figure}[htb!]
\plotone{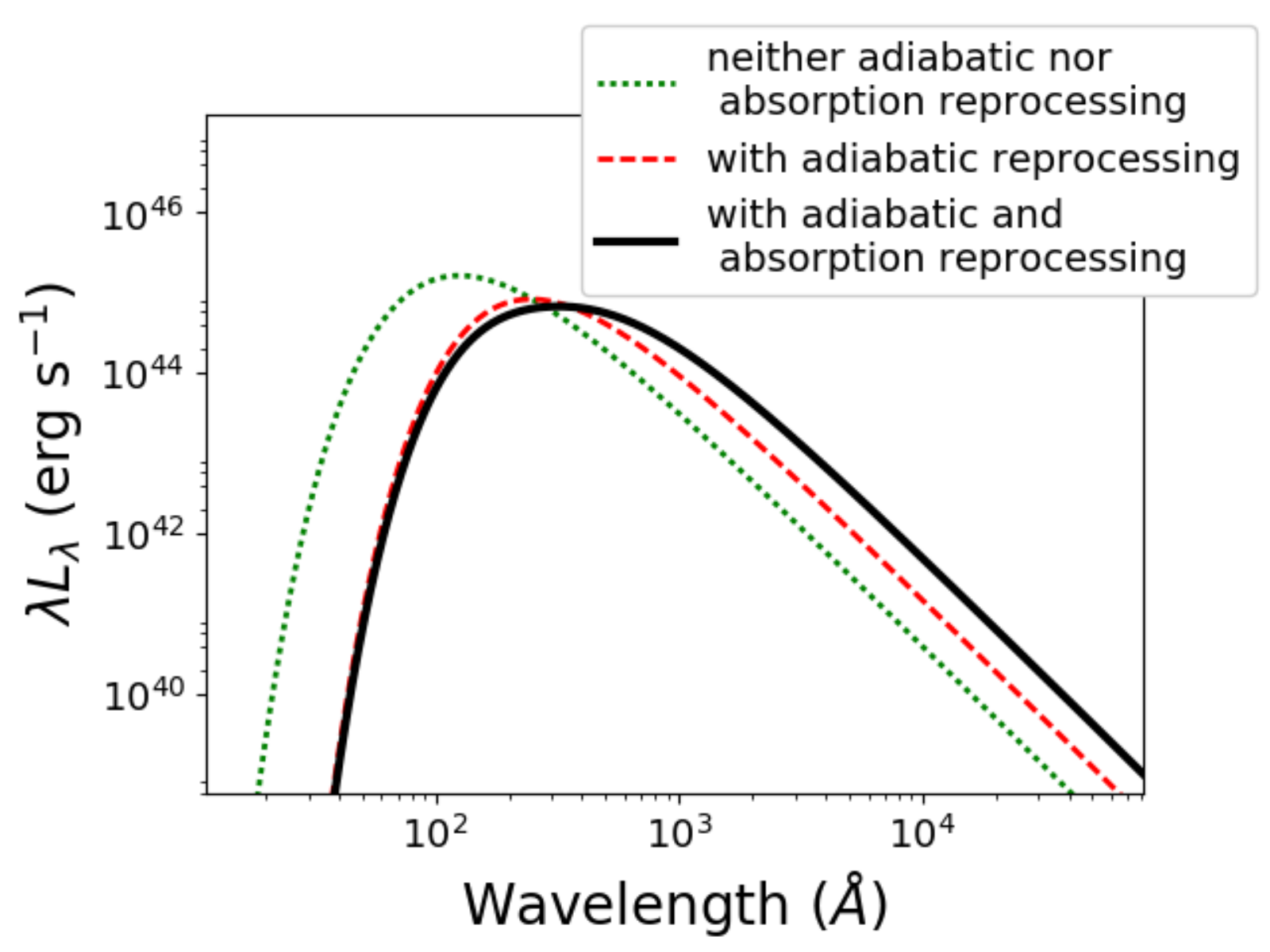}
\caption{Spectral energy distributions from our simplified continuum-reprocessing approach. The dotted green curve shows the thermal emission for $T = 2.93 \times 10^5$ K, corresponding to the static atmosphere model of Appendix~\ref{sec:AdiabaticLossesWithEngine}. The dashed red curve shows the SED when the adiabatic losses and advective effects are included from a homologous outflow with $v_{\rm sc}= 10^4$ km s$^{-1}$. The solid black curve is the result of applying the ``two-temperature'' absorption and re-emission model described in Appendix~\ref{sec:TwoTemperatureReprocessing}.}
\label{fig:SedsReprocesseingStages}
\end{figure}

Figure~\ref{fig:SedsReprocesseingStages} shows spectral energy distributions (SEDs) that summarize the results of our simplified continuum reprocessing model described in this and the previous section. For the results presented in the rest of the paper, we will incorporate the adiabatic and advective effects of an expanding atmosphere along with the ``two-temperature'' absorption and re-emission model, unless we state otherwise.

\subsection{NLTE Solution}
\label{sec:NLTE}

From equation~\eqref{eq:TwoTemperatureReprocessing}, we have an approximate formula for the continuum radiation field at every radius. We can now use this to compute photoionization rates and line fluxes $\bar{J}_{\lambda,\rm line}$, and solve the non-LTE equations assuming statistical equilibrium. We track transitions for H up to principal quantum number 6, and for \ion{He}{2} up to principal quantum number 9. We do not include any other elements in the NLTE solution. We obtain the ionization state and bound-electron level populations for H and He at each radius, which we turn into line opacities and emissivities.

\begin{figure*}[htb!]
\includegraphics[width =\textwidth]{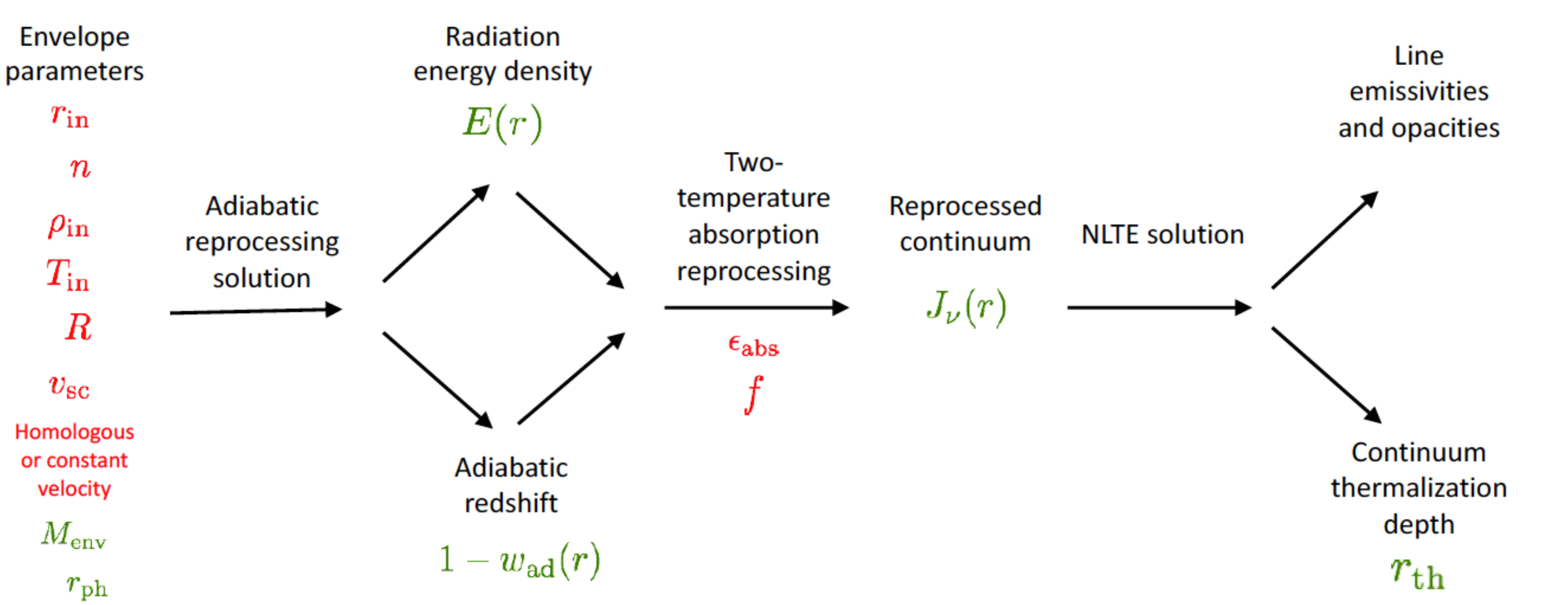}
\caption{Flowchart illustrating the calculation stages needed to turn the input parameters into the final line emissivities and opacities. Input parameters are in red text. Derived values are in green. There are seven envelope parameters, mainly covering uncertainty in the details of the hydrodynamics. These parameters might also be connected to the BH mass, stellar mass, stellar structure, etc, and are also functions of time. We assume a solar abundance ratio of H to He throughout. The two reprocessing parameters, $\epsilon_{\rm abs}$ and $f$, are in principle related to the non-LTE solution, and would require an iterative procedure to find self-consistently. However, we simplify this process by choosing values for them at the outset.}
\label{fig:FlowChart}
\end{figure*}

Figure~\ref{fig:FlowChart} summarizes the entire calculation process up this point (Appendices~\ref{sec:AdiabaticLossesWithEngine} through \ref{sec:NLTE}), reviewing how the envelope input parameters are turned into line emissivities and opacities for the final radiative transfer calculation, from which we find the final line profile.

\subsection{Using the NLTE Results in Radiative Transfer Calculations}
\label{sec:RadiativeTransferSetup}

In the final radiative transfer calculations, we make use of the tabulated line opacities and emissivities described in Appendix~\ref{sec:NLTE}, which set the rate at which photons are emitted and absorbed throughout the calculation volume.

To determine the electron temperature, $T_e$ as a function of position, we use $T_e = \left(E_{\rm rad} / a_{\rm rad}\right)^{1/4}$, where $a_{\rm rad}$ is the radiation constant, and $E_{\rm rad}$ is the radiation energy density solution described in Appendix~\ref{sec:AdiabaticLossesWithEngine}. In reality, $T_e$ is set by radiative equilibrium, accounting for the heating and cooling processes of the electrons as was done in R16. The result is that, compared to R16, we tend to underestimate the electron temperature near the surface of the envelope. On the other hand, R16 did not include the full range of metals that can contribute to electron cooling; if included, they would raise the opacity of the gas and allow it to reach a value closer to the one given by the integrated radiation energy density that we are using in this paper. 

We adjust the inner photosphere $r_{\rm th}$  to correspond to the thermalization depth of the continuum at the line-center wavelength. To calculate the thermalization depth, we assume that free-free processes are dominating the continuum emission. In more detail, we define $r_{\rm th}$ as the radius corresponding to $\tau_e = 1/(\sqrt{3 \epsilon_{\rm ff}})$, where $\epsilon_{\rm ff}$ is the ratio of the free-free opacity to the electron scattering opacity, and we evaluate these opacities based on the density and temperature at the outer edge of the envelope. As we adjust parameters, if we encounter a situation where $r_{\rm th}$ would fall within $r_{\rm in}$, we use $r_{\rm in}$ as the inner boundary.

\section{Treatment of Non-coherent (Compton) Scattering}
\label{Comptonization}

The non-coherence of electron scattering is a combination of three effects: (1) Doppler shifts introduced when boosting between the observer's frame and the initial rest-frame of the electron; (2) the post-scattering recoil of the electron, as measured in its initial rest frame; and (3) the requirement for the photon phase space density to obey Bose-Einstein statistics. Ignoring spatial dependencies, in the limit of many scatterings, and for small enough electron temperatures such that the electrons are non-relativistic, the evolution of the photon phase-space density can be written in the form of a Fokker--Planck equation commonly known as the Kompaneets equation \citep{Kompaneets1957}. For photon frequency $\nu$ and radiation spectral energy density $u_\nu$, when $u_\nu c^3/ (8 \pi h \nu^3) \ll 1$, the third effect and its corresponding terms may be neglected.  This is the case in many astrophysical applications, and we assume it is the case here.

The Kompaneets equation begins to lose accuracy at optical depths of order unity; to account for spatial and temporal variation, it must be combined with the radiative transfer equation. Several highly accurate numerical techniques have been developed to accomplish this \citep[e.g.][]{Rybicki1994}. Here, we use a Monte Carlo treatment of the scattering process. Such an approach has been used for this particular problem many times in the past, starting with \citet{Auer1972} and notably by \citet{Pozdnyakov1983}.

We account for Doppler shifts due to fluid motion by first performing Lorentz transformations to boost into the co-moving frame of the fluid. We similarly account for thermal motion of the electrons by randomly sampling a velocity from a Maxwell-Boltzmann distribution, following the procedure described in \citet{Pozdnyakov1983}, and then boosting into the electron rest frame. We then sample the outgoing photon direction from the classical Thomson differential scattering cross section (the Rayleigh phase function). While straightforward to include, we have omitted Klein-Nishina corrections to the total and differential cross sections, which are negligible for the photon energies of interest to us ($h \nu \ll m_e c^2$). We account for the change in photon energy due to electron recoil. Finally, we apply the inverse Lorentz transformations to move back to the co-moving frame of the fluid, and then back to the lab frame.

We have validated our scattering implementation by performing a one-zone test problem similar to that described in \citet{Castor2004} \citep[see also][]{Ryan2015}, which tests how well we capture both effects 1 and 2 listed above. An initially monochromatic collection of photons interact with a population of thermal electrons solely via scattering. We initialize the electrons at temperature $T_e$, density $n_e$, and zero bulk velocity. We inject photons at initial frequency $\nu_0$, which we nondimensionalize to $x_0$, where $x_0 = h \nu_0 / k_B T_e$, and with a total radiative energy density, $u_{\rm rad}$, that is small compared to that of the integrated electron kinetic energy, so that $T_e$ will remain approximately constant over the duration of the calculation. As the photons scatter, their energy distribution evolves on a timescale $t_c$ given by
\begin{equation}
t_c =  \frac{1}{n_e \sigma_T c} \, \frac{m_e c^2}{4 k_B T_e} 
\end{equation}

In the absence of stimulated scattering, the photon energy spectrum should converge to a Wien distribution with mean intensity $J_{\nu, \,\,{\rm final}}$ given by

\begin{equation}
J_{\nu,\,\,{\rm final}} = \frac{2 h \nu^3}{c^2} \exp\left[-\left(\frac{h \nu}{k_B T_e} + \mu \right)\right]
\end{equation}

where $\mu$ is the chemical potential. Applying conservation of photon number, we find
\begin{equation}
  \label{eq:ChemicalPotential}
  \mu = -\ln \left[ \frac{1}{2} \left(\frac{h}{k_BT_e} \right)^3 \frac{u_{\rm rad}\,c^3}{8 \pi h \nu_0}\right]
\end{equation}

\begin{figure}[htb!]
\plotone{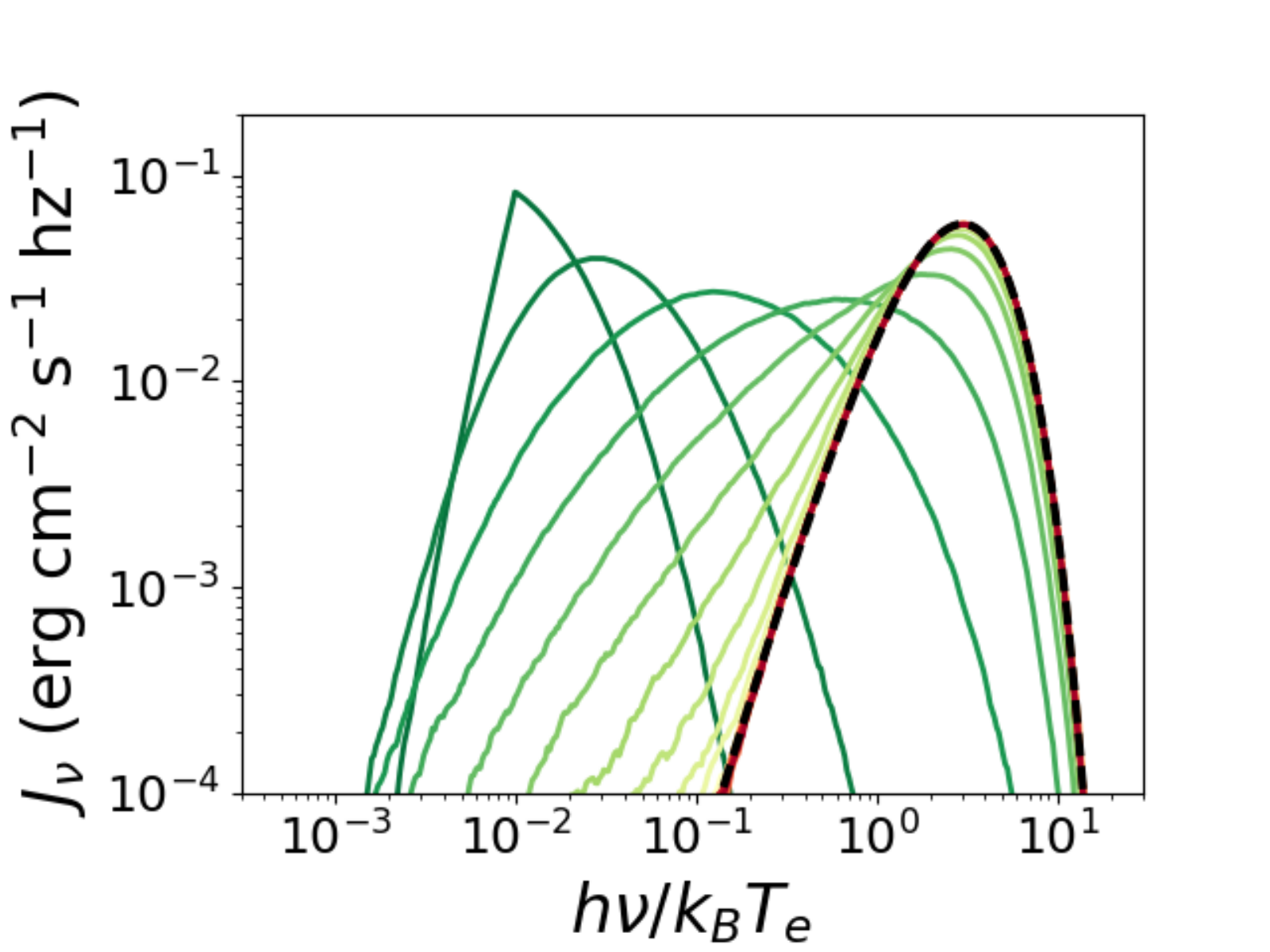}
\caption{A test of the Comptonization physics implemented in the Monte Carlo radiative transfer. An initially monochromatic collection of photons interact with a bath of thermal electrons solely via scattering. The photon energy distribution converges to a Wien distribution (dashed black line) at a rate governed by the initial temperature of the electrons and the number of scatterings undergone by the photons. The curves are separated by time intervals of $2 \,t_c$. The parameter values are $T_e = 10^6$~K, $n_e = 10^{12}$~cm$^{-3}$, $x_0 = 0.01$, and $u_{\rm rad} = 10^{-6} a_{\rm rad} T_e^4$.}
\label{fig:WienTest}
\end{figure}

The results of such a test with $T_e = 10^6$~K, $n_e = 10^{12}$~cm$^{-3}$, $x_0 = 0.01$, and $u_{\rm rad} = 10^{-6} a_{\rm rad} T_e^4$ are shown in Figure~\ref{fig:WienTest}.

\end{document}